\documentclass[prd,twocolumn,aps,letterpaper,amsmath,amssymb,preprintnumbers,showpacs,superscriptaddress,floatfix,dvipsnames,table,longbibliography,nofootinbib]{revtex4-2}
\usepackage[T1]{fontenc}
\usepackage{array}
\usepackage{graphicx}
\usepackage[export]{adjustbox}
\usepackage[caption=false]{subfig}
\usepackage{tikz}
\usetikzlibrary{tikzmark}
\usetikzlibrary{calc}
\allowdisplaybreaks 
\usepackage[hypertexnames=true]{hyperref}
\usepackage[capitalize]{cleveref}
\hypersetup{
    colorlinks=true,       
    linkcolor=blue,          
    citecolor=blue,        
    filecolor=blue,      
    urlcolor=blue           
}

\usepackage{slashed}
\usepackage{mathtools}
\usepackage{listings}
\usepackage{pgf}
\usepackage{fancyvrb}
\usepackage{longtable}
\usepackage{bm}
\usepackage{tikz-cd}
\usepackage{amsthm}
\usepackage{mathrsfs}
\usepackage[normalem]{ulem}
\usepackage[all]{foreign}

\newdimen\defaultaddspace
\DeclareMathOperator{\imag}{Im}
\DeclareMathOperator{\real}{Re}
\DeclareMathOperator{\argmin}{argmin}
\DeclareMathOperator{\sn}{sn}

\newcommand{\R}[0]{\ensuremath{\mathbb{R}}}
\newcommand{\C}[0]{\ensuremath{\mathbb{C}}}
\newcommand{\disk}[0]{\ensuremath{\mathbb{D}}}
\newcommand{\abs}[1]{\ensuremath{| #1|}}

\begin{document}
\title{Bayesian Analysis and Analytic Continuation \texorpdfstring{\\}{} of Scattering Amplitudes from Lattice QCD}
\begin{abstract}
We present a novel procedure for analyzing the lattice-QCD spectrum via the finite-volume formalism to obtain constraints on multi-hadron scattering amplitudes at both real and complex energies.
This approach combines a Bayesian reconstruction of the scattering amplitude on the real axis with Nevanlinna interpolation for analytic continuation to complex-valued energies.
The method is non-parametric, inherently accounting for parametrization dependence within the uncertainty.
We demonstrate the applicability of this approach using both toy data and real lattice-QCD data in resonant systems from the HadSpec and BaSc collaborations.
\end{abstract}

\author{Miguel Salg}
\email{miguel.salg@unibe.ch}
\affiliation{Albert Einstein Center, Institute for Theoretical Physics, University of Bern, 3012 Bern, Switzerland}
\author{Fernando Romero-L\'opez}
\email{fernando.romero-lopez@unibe.ch}
\affiliation{Albert Einstein Center, Institute for Theoretical Physics, University of Bern, 3012 Bern, Switzerland}
\author{William~I.~Jay}
\email{william.jay@colostate.edu}
\affiliation{Department of Physics, Colorado State University}

\date{October 2025}

\maketitle

\section{Introduction}

With the discovery of new unstable hadrons at experimental facilities~\cite{LHCb-FIGURE-2021-001-report, LIU20232148}, the need to describe the hadron spectrum from first-principles quantum chromodynamics (QCD) has become increasingly urgent.
A successful approach involves constraining scattering amplitudes from lattice QCD~\cite{Bulava:2022ovd} by relating the finite-volume energy spectrum of multi-hadron systems in QCD to the scattering amplitude~\cite{Luscher:1986pf}.
The resulting amplitude can be analytically continued to the complex energy plane to determine the masses, widths, and couplings of hadronic resonances. Despite its success, this method faces challenges, particularly the need to extrapolate from the real axis to complex-valued energies.

The standard approach to obtain multi-hadron scattering amplitudes employs the finite-volume formalism, which relates the finite-volume energy spectrum, characterized by a specific set of quantum numbers, to the infinite-volume scattering amplitude.
In the two-body case, each energy level imposes a constraint on the scattering amplitude at the corresponding energy~\cite{Luscher:1986pf,Luscher:1990ux,Rummukainen:1995vs,Kim:2005gf,Bernard:2008ax,Leskovec:2012gb,Gockeler:2012yj,Briceno:2014oea}---see Ref.~\cite{Briceno:2017max} for a review. In practice, multiple energy levels, obtained using a matrix of correlation functions built from several operators, are used as constraints, and simple functional forms are fitted to these data points.
For simple functional forms, like the effective-range expansion, the analytic continuation to complex energies is trivial.
However, the choice of a particular parametric description introduces model dependence and systematic uncertainties that are hard to quantify.
The challenge is particularly keen in certain cases of phenomenological interest, including broad resonances like the $\sigma$, far away from threshold where the effective-range expansion is poorly convergent, or in multi-channel systems~\cite{BaryonScatteringBaSc:2023zvt,BaryonScatteringBaSc:2023ori,Rodas:2023gma,Briceno:2017qmb,Prelovsek:2020eiw,Briceno:2016mjc,Dudek:2014qha,Woss:2019hse,Wilson:2015dqa}.

In this work, we present a novel perspective on the extraction of scattering amplitudes from lattice QCD.
This approach is inspired by viewing the finite-volume formalism as an inverse problem, namely, numerical analytic continuation from a finite set of points.
While a known amplitude can fully reconstruct the spectrum in the elastic region, the available spectrum only provides a finite set of constraints on the amplitude.
Fortunately, unlike other inverse problems such as the reconstruction of spectral densities (see Ref.~\cite{Rothkopf:2022fyo} for a recent review), this problem is not ill-conditioned, which has led to numerous numerical applications in hadron spectroscopy.
This perspective allows us to adapt tools commonly used in inverse problems to the analysis of scattering amplitudes.

Specifically, our work receives inspiration from two different tools.
One is the Bayesian reconstruction of inverse problems using Gaussian processes~\cite{Horak:2021syv,DelDebbio:2021whr,Pawlowski:2022zhh,Horak:2023xfb,DelDebbio:2024lwm,Candido:2024hjt,Dutrieux:2024rem,Dutrieux:2025jed}.
The other is analytic continuation based on Nevanlinna--Pick interpolation~\cite{PhysRevLett.126.056402,Bergamaschi:2023xzx}.
The combination of these two approaches allows us to quantify the uncertainty on scattering amplitudes and resonance poles in a controlled way, including the role of model dependence.

Our analysis method proceeds in two steps.
First, a parameterization-agnostic Bayesian method, inspired by Gaussian processes, constrains the amplitudes for real-valued energies.
Second, Nevanlinna--Pick interpolation leverages the known analytic structure of the amplitude in the complex plane to constrain the location of possible resonance poles.
We argue that the Bayesian uncertainty of the first step incorporates the parameterization dependence, while techniques from Nevanlinna--Pick interpolation rigorously bound the uncertainty from ``analytically continuing'' from a finite set of points, assuming analyticity in the extrapolation region.\footnote{Applying Nevanlinna--Pick interpolation requires mapping the problem to the unit disk using conformal maps.
For the rigorous bounds to apply, the conformal maps must correctly bound the codomain of the inverse scattering amplitude.
\Cref{sec:analytic_continuation} discusses this point in detail.
}

This paper is organized as follows. We start with a brief review of the two-hadron scattering formalism in finite and infinite volume in \cref{sec:formalism}. This is followed by a detailed account of our proposed non-parametric analysis strategy: in \cref{sec:bayes}, we discuss how the Bayesian approach proceeds for the analysis at real energies, while in \cref{sec:analytic_continuation}, we present Nevanlinna--Pick interpolation for the analytic continuation to complex energies. In \cref{sec:toy_example}, a toy example inspired by the $\rho$ resonance is presented. Numerical applications to maximal-isospin $\pi\pi$ scattering, the $\sigma$ resonance, and the $\Lambda(1405)$ resonance are discussed in \cref{sec:analysisqcd}, comparing to existing and published lattice-QCD data by the Hadron Spectrum (HadSpec) and the Baryon Scattering (BaSc) collaborations~\cite{BaryonScatteringBaSc:2023zvt,Rodas:2023gma,Briceno:2017qmb,Briceno:2016mjc}.
We conclude in \cref{sec:conclusion}.
\Cref{app:HMC,app:Bayes_factors_approx} collect a few technical results.

\section{Two-hadron scattering}
\label{sec:formalism}

In this work, we consider two-body scattering, with both single and multiple channels.
This section reviews the relevant formalism both in finite and infinite volume.
It also describes the standard fitting approach used to infer infinite-volume scattering observables from lattice-QCD calculations using the finite-volume formalism.

\subsection{General considerations of two-hadron scattering amplitudes}

\begin{figure*}[t]
    \centering
    \includegraphics[width=0.8\linewidth]{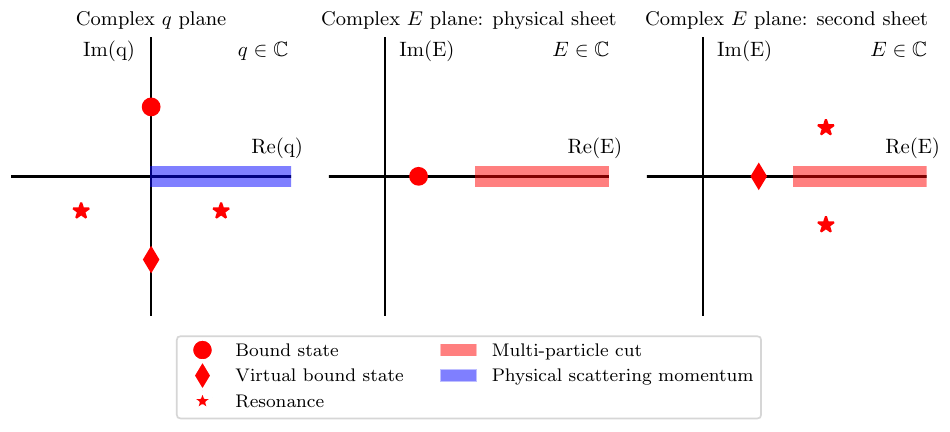}
    \caption{Locations of singularities in the scattering amplitude $\mathcal{M}_\ell$ in the complex momentum and energy planes, inspired by Fig.~1 in Ref.~\cite{Matuschek:2020gqe}.
    }
    \label{fig:scattering_singularities}
\end{figure*}

In the following, we assume spinless particles or other cases where $J^P$ is uniquely determined by $\ell$, such as scattering between pseudoscalar mesons and baryons.
The central object of interest is the partial-wave projected two-hadron scattering amplitude, $\mathcal{M}_\ell$, where $\ell$ denotes the partial wave. $\mathcal M_\ell$ is a function of only the center-of-mass energy of the system $E=\sqrt{s}$. For a specific scattering channel, we suppose that the two scattering particles have masses $m_1$ and $m_2$ so that
\begin{equation}
    \sqrt{s} = \sqrt{m_1^2 + \bm{q}^2} + \sqrt{m_2^2 + \bm{q}^2},
    \label{eq:energy}
\end{equation}
where the magnitude of the momentum in the center-of-mass frame is $q = \abs{\bm{q}}$.

A standard representation of the scattering amplitude is in terms of the partial-wave projected K matrix $\mathcal{K}_\ell$:
\begin{align}
    \mathcal M_\ell^{-1} = \mathcal K_\ell^{-1} - i \rho,
    \label{eq:scatteringAmplitude}
\end{align}
where $\rho= q/(\eta 16 \pi \sqrt{s})$ is the phase space, with $\eta=1$ for identical particles and $\eta = 1/2$ for non-identical particles.
Unitarity and causality dictate that $\mathcal K_\ell$ be a meromorphic function, which is real-valued for real energies above sub-threshold left-hand cuts.
In the single-channel case, the K matrix is related to the phase shift $\delta_\ell$,
\begin{equation}
    \mathcal{K}^{-1}_\ell = \frac{q \cot \delta_\ell}{\eta 16 \pi \sqrt{s}}.
\end{equation}
In other words, the scattering amplitude may be equivalently parameterized in terms of either the phase shift or the K matrix. Around $q^2 \sim 0$, the inverse K matrix can be expanded as
\begin{equation}
    q^{2\ell + 1}\cot \delta_\ell (q^2) = -1/a_\ell + r_\ell q^2/2 + \mathcal{O}(q^4),
    \label{eq:ERE}
\end{equation}
where $a_\ell$ is the scattering length and $r_\ell$ is the effective range; this is referred to as effective-range expansion.  Note, however, that this is a very simple parametrization insufficient to describe  the energy dependence in most cases, since the expansion is only accurate for asymptotically small $q^2$.

Considering the single-channel $\mathcal M_\ell$ as a function of complex energy, the square root relating $q$ and $E$ in \cref{eq:energy} implies that $\mathcal M_\ell$ has a branch cut starting at threshold $E = m_1+m_2$.
The two corresponding Riemann sheets can be classified by the sign of $\imag q$:
the sheet with $\imag q > 0$ is the physical sheet, while the sheet with $\imag q < 0$ is the unphysical sheet.

An important goal in the study of scattering amplitudes is the identification of the types and locations of the poles of $\mathcal{M}_\ell$ because they correspond to hadronic resonances, bound states, or virtual bound states, appearing in the scattering process.
The function $\mathcal{M}_\ell(E)$ cannot have poles off the real energy axis on the physical sheet (corresponding to the upper half plane excluding the positive imaginary axis in $q$), as this would violate causality~\cite{Gribov2023}.
Neither are poles on the real energy axis above threshold allowed (corresponding to the positive real $q$ axis), which would be at odds with unitarity.\footnote{Unitarity implies that $\imag \mathcal{M}_\ell \propto \abs{\mathcal{M}_\ell}^2$ and that \cref{eq:scatteringAmplitude} is fulfilled with both $\mathcal{K}_\ell^{-1}$ and $\rho$ being real for real energies above threshold. These two conditions cannot simultaneously be satisfied at a pole.}
In all other cases, poles are possible and are interpreted as different types of hadrons, depending on the pole location in the complex plane. A pole on the real energy axis below threshold is associated with a bound state if it is on the physical sheet (\ie on the positive imaginary $q$ axis) and with a virtual bound state if it is on the unphysical sheet (\ie on the negative imaginary $q$ axis). Poles off the real energy axis are, as mentioned before, only allowed on the unphysical sheet (\ie in the lower half plane in $q$), and they always appear in complex conjugate pairs. They are interpreted as resonances.
The singularity structure of $\mathcal M_\ell$ in the complex momentum and energy planes is summarized visually in \cref{fig:scattering_singularities}.

In the vicinity of a pole, the scattering amplitude behaves like
\begin{equation}
    \mathcal M_\ell \simeq -\frac{g^2}{E^2 - E_R^2} ,
\end{equation}
where $E_R = M_R -i \Gamma/2$, with the resonance mass $M_R$ and its width $\Gamma$.
For bound states and virtual bound states, $\Gamma = 0$ and $M_R$ is the bound-state mass. For bound states, $(-1)^\ell g^2>0$, as required to generate a normalizable state, while for virtual states it can take any value.
For resonances, $g$ is complex.

Returning to single-channel $s$-wave scattering, substituting \cref{eq:ERE} into
\cref{eq:scatteringAmplitude} yields poles on the imaginary axis at
$q = i/r_0 \left( 1 \pm \sqrt{1-2r_0/a_0}\right)$, which can be either positive $q=i|q|$ or negative $q=-i|q|$. More general parameterizations of the inverse K matrix can produce other pole structures, such as resonance poles displayed in \cref{fig:scattering_singularities}.
In practice, bound states, virtual bound states, and resonances can be distinguished by checking numerically their position in the complex momentum plane.

\Cref{eq:scatteringAmplitude} can also be used to represent multichannel scattering amplitudes.
In this case, $\mathcal{K}_\ell$ is a matrix with channel indices, $\mathcal{K}_{\ell, ab}$, and $\rho_{ab} = \delta_{ab} q_a / (\eta_a 16\pi\sqrt{s})$.

\subsection{Finite-volume formalism}
Next we turn to the finite-volume formalism, the goal of which is to constrain $\mathcal M_\ell$ from the lattice-QCD energy spectrum~\cite{Luscher:1986pf,Luscher:1990ux}. The finite-volume quantization condition~\cite{Luscher:1990ux,Kim:2005gf,Briceno:2014oea}
\begin{align}
    \det_{a \ell m} \left.\left[ \mathcal K^{-1}(E) + F(E, \bm{P}, L)\right] \right|_{E=E_n} = 0,
    \label{eq:QC:general}
\end{align}
provides the connection between finite-volume energy levels $E_n$ determined using lattice QCD and the partial-wave-projected K matrix $\mathcal{K}_\ell$.
Here, $F(E,\bm{P}, L )$  is a matrix whose entries are known functions that depend on kinematical variables, and can be evaluated numerically.
The determinant runs over open channels $a$, partial-wave indices up to a maximal partial wave, $\ell\leq\ell_{\rm max}$, and the integer $m$ which satisfies $-\ell \leq m \leq \ell$.
Both $F$ and $\mathcal{K}$, and thus the whole matrix under the determinant, become block diagonal when expressed in terms of the irreducible representations (irreps) of the relevant finite-volume symmetry group.

In the single-channel case, the quantization condition can be expressed in a simpler form.
Assuming $s$-wave dominance ($\ell_{\rm max} = 0$) and identical particles, the quantization condition \cref{eq:QC:general} takes the form:
\begin{equation}
    q \cot \delta_0 (q^2) = \frac{2}{\gamma L \pi^{1/2}} \mathcal Z_{00}\left( q^2, L, \boldsymbol{P} \right),
    \label{eq:QC2}
\end{equation}
where $\gamma = (s + \boldsymbol{P}^2)^{1/2}/\sqrt{s}$, $\boldsymbol{P}$ is the finite-volume momentum of the system, $L$ is the spatial size of the box, and $\mathcal{Z}_{00}$ is a L\"uscher zeta function.

We now turn to the multichannel case and, for concreteness, focus on the coupled $\pi\Sigma - \bar K N $ system where the $\Lambda(1405)$ is present. Assuming $s$-wave dominance, the quantization condition can be written as
\begin{align}
    \det \left[
    \begin{pmatrix}
        \mathcal K^{-1}_{\pi\Sigma;\pi\Sigma} & \mathcal K^{-1}_{\pi\Sigma;\bar{K}N} \\
        \mathcal K^{-1}_{\bar{K}N;\pi\Sigma} & \mathcal K^{-1}_{\bar{K} N;\bar{K} N}
    \end{pmatrix}
    +
    \begin{pmatrix}
        F_{\pi\Sigma} & 0 \\
        0 & F_{\bar K N}
    \end{pmatrix}
    \right] = 0,
    \label{eq:QC:2channel}
\end{align}
where $F_{\pi\Sigma}$ and $F_{\bar K N}$ are known functions, \ie matrix entries of $F(E, \boldsymbol{P}, L)$ in \cref{eq:QC:general}. Moreover, the K matrix is symmetric, \ie $\mathcal K^{-1}_{\pi\Sigma;\bar{K}N} = \mathcal K^{-1}_{\bar{K}N;\pi\Sigma}$.
For each energy, multichannel unitarity constrains the imaginary parts so that only three real numbers $\mathcal K^{-1}_{\pi\Sigma;\pi\Sigma}$, $\mathcal K^{-1}_{\pi\Sigma;\bar{K}N}$, and $\mathcal K^{-1}_{\bar{K}N;\bar{K}N}$ are needed to describe the scattering amplitude. In other words, each energy gives a single real equation with three unknowns.

\subsection{Standard fitting approach in lattice QCD}

In the single-channel case and assuming $s$-wave dominance, the quantization condition \cref{eq:QC2} can be used to determine the value of the K matrix, or, equivalently, the phase shift, at each energy where a finite-volume level is present. However, solving the quantization condition only yields a discrete set of points.
In order to carry out analytic continuation to the complex energy plane and locate resonance poles, a common approach is to parameterize the K matrix, its inverse, or the phase shift as a continuous function of energy.

For this purpose, one chooses some explicit parametric model for, \eg, $\mathcal K_\ell^{-1}(E; \bm \theta)$, where $\bm \theta$ denotes free parameters of the model which are adjusted by a fit using \cref{eq:QC:general}.
Examples of typically employed models include polynomials or ratios of polynomials.
For a given model, one proceeds by minimizing a $\chi^2$ function relating the lattice-QCD energy levels $\bm E_\mathrm{d}$ and the predicted energy levels $\bm E_\mathrm{QC}(\bm \theta)$ obtained from solving \cref{eq:QC:general} for specific parameter values $\bm \theta$,
\begin{align}
    \label{eq:chi2Fit}
    \chi^2(\bm \theta) &= (\bm E_\mathrm{d} - \bm E_\mathrm{QC}(\bm \theta))^T \Sigma_\mathrm{d}^{-1} (\bm E_\mathrm{d} - \bm E_\mathrm{QC}(\bm \theta)) , \\
    \label{eq:chi2opt}
    \bm \theta_\mathrm{opt} &= \argmin_{\bm \theta} \chi^2(\bm \theta) ,
\end{align}
where $\Sigma_\mathrm{d}$ is the data covariance matrix. For a given set of parameters, the process of finding $\bm{E}_\mathrm{QC}(\bm{\theta})$ is referred to as ``solving the quantization condition.''
Such an analysis induces an intrinsic model dependence as the usual models are rather restrictive to avoid overfitting.
In particular, the pole positions related to resonances will potentially depend on the model chosen.

In the multichannel case or when considering higher partial waves, the situation is further complicated by the absence of a one-to-one relation between energy and the K matrix. As pointed out at the end of the previous subsection, there is still only one constraint at each finite-volume energy, but several unknowns in the infinite volume.
In the following, we will again consider the case of $\Lambda(1405)$ for concreteness.
The standard approach is to choose an explicit parametric model for the unknown $\mathcal K^{-1}_{\pi\Sigma;\pi\Sigma}$, $\mathcal K^{-1}_{\pi\Sigma;\bar{K}N}$, and $\mathcal K^{-1}_{\bar{K}N;\bar{K}N}$ as functions of energy.
The parameters of the model are fixed by using \cref{eq:QC:2channel} to predict the finite-volume energies corresponding to the model and comparing to the energy levels computed from lattice QCD, analogously to \cref{eq:chi2Fit,eq:chi2opt}.

As emphasized in Ref.~\cite{Briceno:2017max}, careful study of the parameterization dependence of results is essential to quantify systematic uncertainties in this approach.
A possibility that has been employed in the literature \cite{Dawid:2025doq,Boyle:2024grr,Boyle:2024hvv} is to perform a model average~\cite{Jay:2020jkz}, but this is still restricted to a finite set of models.
In contrast, the approach pursued in the present work and outlined in detail in the following two sections aims to solve the same problem non-parametrically.
Another approach followed in the literature \cite{Rodas:2023nec} involves using crossing symmetry not only to quantify, but to significantly reduce the parameterization dependence. This method relates amplitudes in different isospin channels and partial waves in a model-independent way, but it is still based on parametric fits. In practice, this method is however only applicable to certain systems, such as two pions.

\section{Bayesian study of scattering amplitudes for real energies}
\label{sec:bayes}

Tools from Bayesian statistics can be employed to place non-parametric constraints on scattering amplitudes at real-valued energies, using lattice-QCD data for the finite-volume spectrum. In this section, we describe in detail how such an analysis can be carried out.

Since no particular functional behavior for $\mathcal M_\ell(E)$ is assumed, a unique analytic continuation to complex energies no longer exists.  Consequently, the methods discussed in this section require additional tools to perform the analytic continuation, which will be addressed in \cref{sec:analytic_continuation}.

\subsection{Bayesian framework}

Given a set of finite-volume energy levels obtained from lattice QCD, the quantization condition can be used to constrain the functional form of $\mathcal K_\ell(E)$.

A general smooth function $u(E): \R \to \R$ can be constructed as an interpolating function between a set of $n_p$ ``nodes'', $\bm E$, and ``values'', $\bm u$. The function can be chosen such that $\mathcal K_\ell^{-1}(\bm E ) = \bm u$, or similar relations up to $q^\ell$ barrier factors.
Note however that the interpolating function is not unique; \cref{sec:analytic_continuation} below describes how to construct such an interpolating function based on Nevanlinna--Pick interpolation. Other approaches, such as splines, are possible (but only for real values of $E$).
Moreover, there is an approximation error induced by having a discrete set, which scales with some power of the spacing between points. In practice, the effects of these ambiguities can be checked by comparing results for different numbers of nodes $n_p$ and/or using different interpolation strategies. In the cases of interest for this work, it has been verified numerically that they are negligible compared to other sources of uncertainty.

In the Bayesian framework, we start with some prior knowledge about the functional form of the K matrix.
This knowledge is encoded in the prior probability density, which we take to be a multivariate Gaussian distribution:
\begin{align}
    \begin{split}
    p(\bm{u} | \bm{u}_0) &= \frac{1}{\sqrt{(2\pi)^{n_p} \det\Sigma_p}} \times \\
    &\qquad \exp\left[ - \frac{1}{2}(\bm{u} - \bm{u}_0 )^T \Sigma_p^{-1}  (\bm{u} - \bm{u}_0) \right].
    \end{split}
    \label{eq:priorProbability}
\end{align}
Here, the central values $\bm{u}_0$ indicate the {\it a priori} expected values of the K matrix.
The covariance matrix $\Sigma_p$ is a functional kernel describing the correlation between points, with entries given by
\begin{equation}
    \Sigma_p(E, E') = \sigma(E,E')^2 \exp \left[ -\frac{(E-E')^2}{2\ell_c^2}\right] + \epsilon \delta_{E,E'}.
    \label{eq:pprior}
\end{equation}
This form of the kernel is known as radial basis function (RBF) kernel, and is a universal kernel that can describe any continuous function~\cite{10.1162/153244302760185252}.
Note that this choice of prior includes several hyperparameters:
the ``correlation length'' $\ell_c$ , which describes the smoothness of the function; $\sigma(E,E')$, often taken to be a constant, which sets the width of the prior; and $\epsilon$, a small number introduced to regulate the covariance matrix.

The values of $\ell_c$ and $\sigma$ are related to the assumptions in the analysis, and ideally, they should be inspired by prior knowledge, \eg previous calculations or other theory constraints. Moreover, while reducing $\ell_c$ decreases the effect of the prior assumptions, the limit $\ell_c \to 0$ cannot be taken due to the finite number of data points. In order to constrain the behavior of the function $u(E)$ between the available data points, it is necessary to set $\ell_c \sim \Delta E$, where $\Delta E$ denotes the typical energy spacing between adjacent points in the vector of nodes $\bm E$.\footnote{To clarify the notation, $\Delta E$ here is unrelated to any physical energy gaps in the system, but merely denotes a hyperparameter of the analysis related to the artificially introduced discretization of energy.} Importantly, results driven by the information in the data should be stable under variations of $\ell_c$ satisfying $\ell_c \sim \Delta E$. Similar considerations apply to the choice of $\sigma$ which should roughly encompass the uncertainty in the data.

The $n_{\rm d}$ lattice-QCD energy levels, $ \bm{E}_\mathrm{d}$, are treated as data in the inference problem
to gain information about the form of $\mathcal K_\ell^{-1}(E)$, where the subscript $\rm{d}$ denotes ``data''.
The likelihood function for the $\bm{E}_\mathrm{d}$ given ``parameters'' $\bm{u}$ is
\begin{align}
    p(\bm{E}_{\rm d} | \bm{u}) &\propto \exp\left[-\frac{1}{2}\chi^2(\bm{u})\right], \label{eq:dataProbability}\\
    \chi^2(\bm{u}) &=\left(\bm{E}_{\rm d} - \bm{E}_{\rm QC}(\bm{u}) \right)^T \Sigma_\mathrm{d}^{-1} (\bm{E}_{\rm d} - \bm{E}_{\rm QC}(\bm{u})), \label{eq:chi2Bayes}
\end{align}
where $\Sigma_{\rm{d}}$ is the data covariance matrix, and $\bm{E}_{\rm QC}(\bm{u})$ are the predicted energy levels from solving the quantization condition given a K matrix constructed as an interpolating function using a set of values $\bm{u}$.

The (unnormalized) posterior density takes the usual form, via Bayes' theorem, of the product of the likelihood with the prior:
\begin{equation}
        p(\bm{u} | \bm{E}_\mathrm{d}, \bm{u}_0) =
        p(\bm{E}_\mathrm{d} | \bm{u})
        \,
        p(\bm{u} | \bm{u}_0).
        \label{eq:posteriorProbability}
\end{equation}
Given $ p(\bm{u} | \bm{E}_\mathrm{d}, \bm{u}_0)$, it is possible to compute expectation values.
In particular, it is possible to assign a central value using
\begin{equation}
    \langle \bm{u} \rangle_{\rm post} = \frac{1}{Z} \int d^{n_p}u \, \bm{u} \,  p(\bm{u} | \bm{E}_\mathrm{d}, \bm{u}_0),
    \label{eq:expectedu}
\end{equation}
where $Z$ is the normalization constant of the posterior distribution. Under the assumption of symmetric errors, the uncertainty of the K matrix is
\begin{equation}
     (\delta \bm{u})^2 = \langle \bm{u}^2  \rangle_{\rm post} - \langle \bm{u}  \rangle_{\rm post}^2.
     \label{eq:erroru}
\end{equation}
Estimation of asymmetric errors is also possible using quantiles.

Note that this procedure is very similar to a Gaussian process~\cite{Horak:2021syv,DelDebbio:2021whr,Pawlowski:2022zhh,Horak:2023xfb,DelDebbio:2024lwm,Candido:2024hjt,Dutrieux:2024rem,Dutrieux:2025jed}.
The key difference is that $E_{\rm QC}(\bm{u})$ is nonlinear in $\bm{u}$, so the usual closed-form expressions for Gaussian processes do not hold here.
In fact, non-linearities often cannot be neglected, due to the singularities present in the zeta function of the quantization condition, \cref{eq:QC2}.

\subsection{Sampling the posterior}
\label{sec:posteriorSampling}

Calculation of the expectation values in \cref{eq:expectedu,eq:erroru} is the crucial step in the Bayesian analysis of the K matrix.
Given the high dimensionality of the relevant integrals, a Monte Carlo approach proves convenient compared to direct quadrature methods.

Sampling from the prior \cref{eq:pprior} is trivial, since the distribution is Gaussian.
This observation suggests the use of reweighting to evaluate posterior expectation values via:
\begin{equation}
    \langle \bm{u} \rangle_{\rm post} = \langle \bm{u} w_{\rm post} \rangle_{\rm prior}, \quad w_{\rm post} = \frac{p(\bm{u} | \bm{ E}_\mathrm{d}, \bm{u}_0)}{p(\bm{u} | \bm{u}_0)}.
    \label{eq:reweightingNaive}
\end{equation}
This method can be very inefficient in practice if the prior and posterior distributions differ significantly.

An improved reweighting approach can be obtained by linearizing the quantization condition. In particular, it is possible to choose a ``linearization point''
$\bm{u}_L$
with
$\bm{E}_L = \bm{E}_{\rm QC}(\bm{u}_L)$,
such that:
\begin{equation}
    \bm{E}_{\rm QC}(\bm{u}) - \bm{E}_L \simeq J (\bm{u} - \bm{u}_L),
    \label{eq:linearizationQC}
\end{equation}
where $J$ is a $n_{\rm d} \times n_p$ matrix, whose entries can be evaluated numerically by a finite-differences approximation.
The linearization of the quantization condition is expected to work well far away from the singularities of the zeta function.
In this way, an approximation of the posterior probability distribution is given by:
\begin{equation}
   p^* = \frac{\exp\left[ - \frac{1}{2}(\bm{u} - \bm{u}^* )^T (\Sigma^*)^{-1} (\bm{u} - \bm{u}^*) \right]}{\sqrt{(2\pi)^{n_p} \det\Sigma^*}},
   \label{eq:approximateProbability}
\end{equation}
where (see also Ref.~\cite{DelDebbio:2021whr})
\begin{align}
    (\Sigma^*)^{-1} &= J^T {\Sigma_{\rm d}^{-1}} J + \Sigma_p^{-1}, \\
     \bm{u}^* &=  \Sigma^* [\Sigma_p^{-1} \bm{u}_0 + J^T {\Sigma_{\rm d}^{-1}} (\bm{E}_\mathrm{d} - \bm{E}_L + J \bm{u}_L)].
\end{align}

The quantities $\Sigma^*$ and $\bm{u}^*$ correspond to the covariance and the expected value of $\bm u$ in the linear approximation.
In practice, one can start with the initial guess $\bm{u}_L = \bm{u}_0$, and then iteratively compute new values for $\bm{u}^*$ and then $\bm{u}_L$ until suitable convergence is obtained.

Using samples drawn from the approximate distribution $ p^{*}$, one can compute expectation values using reweighting as:
\begin{equation}
    \langle \bm{u} \rangle_{\rm post} = \langle \bm{u} w_{*} \rangle_{*}, \quad w_{*} = \frac{p(\bm{u} | \bm{E}_\mathrm{d}, \bm{u}_0)}{p^*(\bm{u} | \bm{E}_\mathrm{d}, \bm{u}_0)},
    \label{eq:reweightingImproved}
\end{equation}
where $\langle \, \cdot \,\rangle_*$ indicates expectation value over samples drawn from $p^*$. This leads to a significant increase in the sampling efficiency compared to reweighting from the prior distribution (\cf \cref{eq:reweightingNaive}), as $p^*$ approximates the actual posterior distribution, thereby accounting for the shift from the prior induced by the data.

This Monte-Carlo evaluation of the integral in \cref{eq:expectedu} introduces a statistical error in the posterior which can be estimated by the usual bootstrap procedure. We remark that since we directly draw our samples from a Gaussian distribution according to \cref{eq:approximateProbability} rather than generating them from a Markov process, autocorrelations are absent here. Using $\mathcal{O}(100)$ samples, the Monte-Carlo integration uncertainty is usually completely negligible compared to the uncertainty from the width of the posterior distribution as per \cref{eq:erroru}.

In certain cases, the posterior distribution can differ significantly from $p^*$, and this sampling procedure is not useful anymore. This can be detected if the reweighting factors have large fluctuations.
In this paper, we have observed such a breakdown of the reweighting approach only in the coupled-channel $\Lambda(1405)$ example (\cf \cref{sec:L1405}), not in any of the single-channel analyses.
A robust approach for sampling in such situations is the Hybrid Monte Carlo (HMC) algorithm, see \cref{app:HMC}. With HMC, however, it becomes difficult to estimate the normalization of the posterior, which is needed for comparing different choices of priors, as explained below.

\subsection{Dependence on prior hyperparameters}
\label{sec:prior_dependence}

The result of the Bayesian analysis depends, in principle, on the prior central values $\bm{u}_0$ as well as on the kernel $\Sigma_p$ and its hyperparameters, \cf \cref{eq:pprior}.
To avoid a large influence of the prior on the final results, the prior width $\sigma$ is typically chosen to be large.
It is important to compare different choices for $\bm{u}_0$ and the kernel hyperparameters to ensure that the level of prior dependence remains acceptable.

In the Bayesian framework, the probability for the model $M^{(1)}$ with prior $\bm{u}_0^{(1)}$ to describe the data $\bm{E}_\mathrm{d}$ over that for the model $M^{(2)}$ with prior $\bm{ u}_0^{(2)}$ is given by,
\begin{equation}
    \frac{p(M^{(1)} | \bm{E}_\mathrm{d})}{p(M^{(2)} | \bm{E}_\mathrm{d})} = \frac{p(\bm{E}_\mathrm{d} | M^{(1)})}{p(\bm{E}_\mathrm{d} | M^{(2)})}
    \frac{p(M^{(1)})}{p(M^{(2)})} .
\end{equation}
The first fraction on the right-hand side is known as the Bayes factor \cite{Jeffreys1935,Kass1995},
\begin{equation}
    B_{12} = \frac{p(\bm{E}_\mathrm{d} | M^{(1)})}{p(\bm{E}_\mathrm{d} | M^{(2)})} .
    \label{eq:BayesFactor}
\end{equation}
The required probabilities are obtained by marginalization over $\bm{u}$,
\begin{align}
    p(\bm{E}_\mathrm{d} | M^{(k)}) &= \int d^{n_p}u \, p(\bm{E}_\mathrm{d} | \bm{u}) p(\bm{u} | \bm{u}_0^{(k)}) \nonumber \\
    &= \int d^{n_p}u \, p(\bm{u} | \bm{E}_\mathrm{d}, \bm{u}_0^{(k)}) = Z_\mathrm{post}^{(k)} ,
    \label{eq:BayesFactorIntegralAsZ}
\end{align}
where we have used \cref{eq:posteriorProbability}.
In other words, the Bayes factor is given by the ratio of the normalization constants of the two posterior probability distributions derived from the two models we aim to compare.

The integral in \cref{eq:BayesFactorIntegralAsZ} can be estimated by a Monte Carlo approach using samples drawn from the approximate probability distribution introduced in \cref{eq:approximateProbability}.
Comparison of \cref{eq:BayesFactorIntegralAsZ,eq:reweightingImproved}
shows that
\begin{align}
Z_\mathrm{post} = \langle w_* \rangle_{*}, \label{eq:Zpost}
\end{align}
\ie the posterior normalization factor equals the average reweighting factor.

Computing $Z_\mathrm{post}$ using \cref{eq:Zpost} requires the reweighting factors to be well defined.
In particular, the prior density \cref{eq:priorProbability}
and the approximate posterior density \cref{eq:approximateProbability} must be normalized.
Since both distributions are Gaussian in the present treatment, the relevant normalization factors are trivial to compute.
The normalization of the likelihood function~\cref{eq:dataProbability} would be more involved due to the predicted energies from the quantization condition.
Fortunately, the normalization is independent of the prior and so cancels in the ratio \cref{eq:BayesFactor}. Thus, explicit normalization of the likelihood is not required.

Comparing the results and Bayes factors associated with different priors allows one to study not only how the final results change, but also to identify priors which lie close to values favored by the data. Barring strong theoretical guidance to the contrary, a desirable prior is one which roughly encompasses all values which are plausible based on the data without constraining them too tightly. The Bayes factors provide a means to render this discussion quantitative.

An important remark concerns the interpretation of the Bayes factors. In particular, the Bayes factor can only be used to compare functional forms of the priors at fixed correlation length and prior width, and not to select the values of $\ell_c$ and $\sigma$.
This is because the Bayes factor has a maximum at $\sigma \to 0$ or $\ell_c \to \infty$, which is a manifestation of the Jeffreys–Lindley paradox~\cite{jeffreys1998theory,Cousins:2013hry,Lindley:1957aa}.
Therefore, the values of $\sigma$ and $\ell_c$ should instead be chosen as discussed below \cref{eq:pprior}.

Finally, we remark that the dependence on the prior and its hyperparameters in the Bayesian approach is typically much milder than the parameterization dependence in the fitting approach.
This will be demonstrated in several numerical examples below, applying the Bayesian approach both to toy data (\cf \cref{sec:toy_example}) and to actual lattice-QCD data (\cf \cref{sec:analysisqcd}).
Moreover, from the Bayesian point of view, it is natural to utilize physics-informed prior knowledge.

\section{ Analytic continuation of scattering amplitudes }
\label{sec:analytic_continuation}

In general, some ambiguity exists in defining the meaning of analytic continuation from a finite set of points.
Many methods employed in the literature \cite{Pelaez:2022qby}, such as continued fractions or Padé approximants, attempt to solve the analytic continuation problem, but do not naturally account for this uncertainty.
Ref.~\cite{Bergamaschi:2023xzx} showed that techniques from Nevanlinna--Pick interpolation naturally and rigorously quantify this ambiguity.

Nevanlinna--Pick interpolation is a method for constructing complex-analytic functions that interpolate certain points~\cite{Pick1915,Nevalinna1919,Nevalinna1929,Nicolau2016}.
The basic inputs are a set of interpolation nodes $\{z_n\}$ and values $\{G_n\}$ with $n\in{1,2,\dots, N}$.
One seeks the set of complex analytic functions satisfying the interpolation condition $G(\{z_n\}) = \{G_n\}$ and obeying physical constraints on the singularity structure of $G$.
The application of ideas from Nevanlinna--Pick interpolation to problems of analytic continuation in field theory was first recognized in Ref.~\cite{PhysRevLett.126.056402}.

\subsection{Review of Nevanlinna--Pick interpolation theory}

The general procedure is the following.
The first step is to transform the problem to the unit disk,
\begin{align}
\begin{split}
    z_n &\mapsto \zeta_n \in \disk,\\
    G_n &\mapsto w_n \in \disk,
\end{split}
\end{align}
seeking a function $f:\disk\to\disk$ which is analytic everywhere on the disk and satisfies the interpolation condition $f(\zeta_n)=w_n$.
Concrete maps taking the original points to the disk are discussed below in \cref{sec:schwarz_christoffel}.

Given pairs of nodes and values $\{(\zeta_n, w_n)\}$, one applies the Schur algorithm \cite{Schur:1918} to enforce the interpolation condition by factoring out suitable Blaschke factors \cite{Blaschke:1915,Garcia:2018},
\begin{equation}
    b_a(\zeta) = \frac{|a|}{a} \frac{a-\zeta}{1-\bar{a}\zeta}, \quad a \in \disk \setminus \{0\} ,
\end{equation}
where $b_0(\zeta) = \zeta$, and the bar denotes complex conjugation.
This process yields a recursive formula for the interpolating function of the form:
\begin{align}
    f(\zeta) = \frac{P_N(\zeta) f_N(\zeta) + Q_N(\zeta)}{R_N(\zeta) f_N(\zeta) + S_N(\zeta)}, \label{eq:nevanlinna-pick-interpolant}
\end{align}
where we recall that $N$ labels the total number of interpolation nodes.
The Nevanlinna coefficients $P_N$, $Q_N$, $R_N$, and $S_N$ are calculable functions of the input data.
Compact recursive formulae for the Nevanlinna coefficients are given in the literature~\cite{Bergamaschi:2023xzx,Nicolau2016}, \eg in Section~V~A of Ref.~\cite{Bergamaschi:2023xzx}.
The basic idea is that the interpolating function becomes more constrained with each addition of a new data point.

The function $f_N: \disk \to \disk$ in \cref{eq:nevanlinna-pick-interpolant} is an arbitrary analytic function, which represents the freedom to include more input data to constrain the interpolating function $f(\zeta)$ further.
By construction, \cref{eq:nevanlinna-pick-interpolant} satisfies the interpolation condition, $f(\{\zeta_n\}) = \{w_n\}$, for any function $f_N: \disk \to \disk$.
When evaluated away from the nodes, suitable interpretation of \cref{eq:nevanlinna-pick-interpolant} furnishes the desired analytic continuation.
For a given fixed $\zeta\in \disk$, varying the possible values of $f_N(\zeta)$ shows~\cite{Nicolau2016,Bergamaschi:2023xzx} that the analytic continuation is constrained to lie within the \emph{Wertevorrat}, a disk $\Delta_N(\zeta) \subset \disk$ with known center and radius.
The Wertevorrat is calculable for any $\zeta \in \disk$ in terms of the input data $\{\zeta_n, w_n\}$ and sharply quantifies the ambiguity in analytic continuation from a finite set of points.
When the Wertevorrat is small, so also is the ambiguity in the analytic continuation.

Recent applications of Nevanlinna--Pick interpolation have focused on the difficult inverse problem of spectral reconstruction~\cite{Bergamaschi:2023xzx,PhysRevLett.126.056402,PhysRevB.104.165111,Nogaki:2023mut}.
Within the present context, the demands of the analytic continuation are milder.
Given discrete samples on the real line for $\mathcal{K}_\ell^{-1}(E)$, one seeks to compute the analytic continuation to the lower half-plane to locate nearby a zero $z_0$ such that $\mathcal{M}^{-1}_\ell(z_0) = 0$.
Since the function $\mathcal{K}_{\ell}^{-1}(z)$ is presumed to be analytic in a neighborhood of the real line and the point $z_0$, a few interpolation points may reasonably be expected to constrain tightly the possible values for the zero $z_0$.
It is important to note that we are applying Nevanlinna--Pick interpolation to each sample from the posterior distribution independently, such that the function values at the interpolation points are known with numerical precision. Therefore, it is still possible to obtain a wide spread of values for $z_0$ in the final result due to the width of the posterior distribution.

\subsection{General considerations for conformal maps \label{sec:mappings_general}}

By construction, solution to the Nevanlinna interpolation problem yields an analytic function $f:\disk \to \disk$.
After mapping the result back to its original domain $U$ and codomain $V$, the solution amounts to an analytic function $\tilde{G}: U \to V$.
If the true function $G$ takes values outside of $V$ for inputs in the domain $U$, \ie, $G(U) \in \widetilde{V} \supset V$,
then the true function will lie outside the solution space of the interpolating function.
In such a case, $\tilde{G}$ clearly yields no rigorous information regarding the true function $G$.
Therefore, a critical requirement is to choose the initial codomain $V$ to be sufficiently large to ensure that the true function $G$ actually belongs to the solution space of the interpolating function.

One can estimate the size of the true codomain $V$ using a fit for a given functional sample.
For instance, fitting a sample $u(E)$ to a polynomial and evaluating the result over the chosen domain $U$ provides an estimate of the bounds on the codomain.
In practice, modest inflation\footnote{In the authors' experience, the precise size of the inflation does not significantly influence the results for the pole locations; inflation of $5\%-15\%$ is used.} of the extremal values for the real and imaginary parts furnishes a rectangular codomain expected to contain $V$.
This method is motivated by the fact that $\mathcal{K}_\ell^{-1}(E)$ is assumed to be free of singularities, and therefore slowly varying, in the region of analytic continuation.

This assumption is one of the major restrictions of the proposed analysis method. In case $\mathcal K^{-1}_\ell(E)$ has a simple pole, but no zeros within the specified region of interest $U$ that is mapped to the unit disk, it is possible to apply Nevanlinna interpolation to its inverse, \ie $\mathcal K_\ell$ itself, and proceed as usual. In case $\mathcal K^{-1}_\ell(E)$ has both poles and zeros, one can employ transformations of the form $f(E) = 1 / (a + \mathcal K^{-1}_\ell(E))$. By tuning the parameter $a$ such that $K^{-1}_\ell(E) \neq -a \; \forall E \in U$, one can achieve that $f(E)$ is analytic in $U$ and thus amenable to Nevanlinna interpolation.
In cases where such a value $a$ does not exist, as is expected in the presence of Castillejo-Dalitz-Dyson (CDD) poles \cite{PhysRev.101.453}, our method is not directly applicable.\footnote{CDD poles are zeros of the scattering amplitude corresponding to poles in $\mathcal{K}_\ell^{-1}$. They are, for instance, required to occur between two poles of the scattering amplitude in a single-channel system. We remark that in a coupled-channel system with two poles of $\mathcal{M}_\ell$, like the example we will discuss in \cref{sec:L1405} below, only one matrix entry of $\mathcal{M}_\ell$ needs to have a zero, so that none of the matrix entries of $\mathcal{K}_\ell^{-1}$ are required to have a pole.}

It is important to emphasize that these considerations only apply to the energy domain considered for the interpolation / analytic continuation. Consequently, the analytic properties of the function outside the considered domain remain unrestricted.

\subsection{Schwarz--Christoffel mappings \label{sec:schwarz_christoffel}}

\begin{figure*}[t!]
    \centering
    \includegraphics[width=0.9\linewidth]{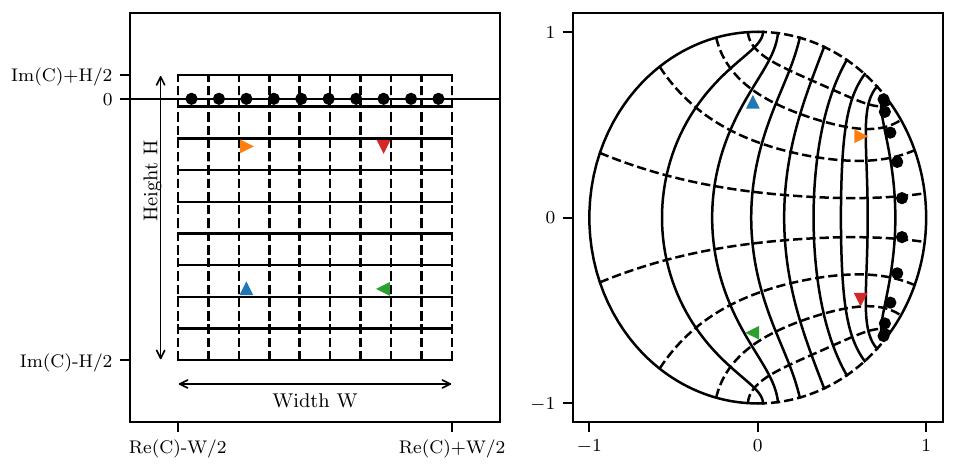}
    \caption{The Schwarz--Christoffel mapping, \cref{eq:schwarz_christoffel_rectangle}, of a rectangle (left panel) to the unit disk (right panel).
    A rectangle of width $W$ and height $H$ is centered at point $C$ below the real line and contains a subset of the real line.
    Black circles denote the schematic locations of input energies on the real line for the analytic continuation problem.
    The colored triangles fix the relative orientation of arbitrary points in the rectangle under the mapping to the unit disk, \eg, the upward-pointing blue triangle on the left is mapped to the upward-pointing blue triangle on the right.
    }
    \label{fig:schwarz_christoffel_rectangle}
\end{figure*}

As guaranteed by the Riemann mapping theorem~\cite{Palka},
there exist conformal mappings that transform an arbitrary simply connected subset of the complex plane to the unit disk.
The Schwarz--Christoffel mapping of a rectangle to the disk is an example of this general result, which is particularly useful for the present application.\footnote{
This conformal map is by no means the only possible choice.
For instance, one could map the ribbon $z\in \R\times(-i,i)$ along the real line to the disk using $f(z) = (e^{\pi z/2}-1)/(e^{\pi z/2}+1)$.
}
Let $z$ denote a point within the rectangle centered at $C \in \C$ with full height (parallel to the imaginary axis) $H \in \R$ and full width (parallel to the real axis) $W \in \R$.
The Schwarz--Christoffel mapping to $\zeta \in \disk$ is then given by \cite{Liu2017,ComputationalConformalMapping}
\begin{equation}
    \zeta = \frac{s - i}{s + i} , \quad s = \sn\left[\left(z - C + i\frac{H}{2}\right) \frac{2K(k)}{W}, k\right],
    \label{eq:schwarz_christoffel_rectangle}
\end{equation}
where $\sn$ denotes a Jacobi elliptic function and $K(k)$ the complete elliptic integral of the first kind of parameter $k$.\footnote{Our notations agree with the ones used in Ref.~\cite{Abramowitz:1972}.}
The value of the parameter $k$ is related to the dimensions of the rectangle via $K(k)/K(\sqrt{1-k^2})=W/(2H)$.
The inverse mapping is given by \cite{Liu2017}
\begin{equation}
    z = F\left[\arcsin\left(i\frac{1 + \zeta}{1 - \zeta}\right), k\right] \frac{W}{2K(k)} - i\frac{H}{2} + C ,
\end{equation}
where $F$ denotes the incomplete elliptic integral of the first kind and $k$ is as above.
It is worthwhile to note that in parts of the literature, $m = k^2$ is used instead of $k$ as parameter for the elliptic functions and integrals, so some care is required with notation.\footnote{This alternative notation is also used in the interface to popular numerical packages including \texttt{scipy} and \texttt{mpmath}.}

The rectangle in energy which is mapped to the unit disk is chosen such that along the real axis, it is centered with respect to our interpolation nodes and extends somewhat beyond them on either end.
Along the imaginary axis, we generally shift the rectangle slightly towards the negative as we intend to find zeros of $\mathcal{M}_\ell^{-1}$ with negative (or zero) imaginary parts, and make sure that the real line is still well contained within the rectangle.
The Schwarz reflection principle ensures that the upper half plane is trivially related to the lower one, so that finding zeros of $\mathcal{M}_\ell^{-1}$ in the lower half plane is sufficient.
A schematic diagram of the situation appears in~\cref{fig:schwarz_christoffel_rectangle}.

The choice of rectangle in the codomain, \ie for the values of $\mathcal{K}_\ell^{-1}$, is detailed in \cref{sec:mappings_general}.
It bears emphasizing that we are parameterizing $\mathcal{K}_\ell^{-1}$ rather than $\mathcal{M}_\ell^{-1}$, so that the target function has no branch point at threshold and is analytic away from sub-threshold left-hand cuts and zeros of $\mathcal{K}_\ell$.\footnote{Regarding cases with zeros of $\mathcal{K}_\ell$ within the specified domain, we refer to the discussion at the end of \cref{sec:mappings_general}.}

\subsection{Numerical considerations}

Numerical calculations with Nevanlinna--Pick interpolation are numerically delicate, in general requiring extended-precision arithmetic for intermediate steps.
As the number of interpolation points grows, numerical cancellations become more severe, and the precision of the input data themselves can become a concern.

Pick showed that the input data must satisfy an analytic self-consistency condition~\cite{Pick1915,Nicolau2016}: the so-called Pick matrix
\begin{align}
    P_{ij} = \frac{1-w_i \bar{w}_j}{1 - \zeta_i\bar{\zeta}_j}
\end{align}
(where the bar denotes complex conjugation) must be positive semi-definite, \ie, all eigenvalues are non-negative so that $\det P \geq 0$.
This condition is known as the Pick criterion, and its practical utility has been emphasized~\cite{PhysRevLett.126.056402}.

For a sufficiently large number of input interpolation points specified with finite precision (say, machine precision of $\approx 10^{-16}$) the Pick criterion may eventually fail.
When the input precision is insufficient, apart from the Pick matrix developing negative eigenvalues, the Wertevorrat may exhibit pathological behavior.
For instance, the Wertevorrat at a given point may saturate or even increase as more data are included in the interpolation problem.

\subsection{Uncertainty analysis \label{sec:pole_locations_Nevanlinna_uncertainty}}

The main goal of the analytic continuation problem is to determine whether $\mathcal{M}_\ell^{-1}(E)$ has zeros, how many such zeros exist, and, if so, to find their locations $E_0$.
The present work incorporates the systematic uncertainty in this quantity using the Wertevorrat in the following manner.

Determining the zeros of $\mathcal{M}_\ell^{-1}(E)$ in this formalism presents an opportunity but also a challenge.
The fundamental complication is that the interpolating function for $\mathcal{K}^{-1}$ is multi-valued when evaluated away from interpolation nodes, since each extrapolation point $\zeta\in\disk$ is mapped to the Wertevorrat,\footnote{Note that $c_N(\zeta)$ and $r_N(\zeta)$ are the center and radius of the Wertevorrat at the point $\zeta$.
Refs.~\cite{Nicolau2016,Bergamaschi:2023xzx} give compact formulas for $c_N(\zeta)$ and $r_N(\zeta)$ in terms of the Nevanlinna coefficients (\cf \eg Eqs.~(56) to (58) in Ref.~\cite{Bergamaschi:2023xzx}).
}
\begin{align}
    \Delta_N(\zeta) = \left\{ c_N(\zeta) + r_N(\zeta) re^{i\theta} : r\in [0,1), \theta\in[0,2\pi] \right\}.
\end{align}
One could imagine using the center of the Wertevorrat as an estimator for the function value when solving $\mathcal{M}_\ell^{-1}(E_0) = 0$ for $E_0$. Instead, the present method uses a random sampling procedure.

For fixed extrapolation point $\zeta\in\disk$, a random point in the Wertevorrat can be sampled by choosing $(r,\theta)$ uniformly within the unit disk
\begin{align}
    (r,\theta) \sim \mathcal{U}_A(\mathbb{D}),
\end{align}
where the right-hand side denotes a uniform distribution with respect to the normalized area measure for the unit disk, $p(r,\theta) =r/\pi$.
In this way, each fixed $(r,\theta)$ yields a single-valued interpolating function
\begin{align}
    f_{(r,\theta)} &: \disk \to \disk \\
    f_{(r,\theta)}(\zeta) &= c_N(\zeta) + r_N(\zeta) re^{i\theta}.
\end{align}
Mapping the domain and codomain back to the original energy coordinates gives a single-valued interpolating function associated with the sampled values for $(r,\theta)$.
This single-valued function will be denoted as $\mathcal{K}^{-1}_{N,(r,\theta)}(E)$.

For a particular sample of the posterior, the values of the random number $(r,\theta)$ are drawn once and held fixed to define $\mathcal{K}^{-1}_{N,(r,\theta)}(E)$ on that sample.
New random points are drawn for subsequent samples of the posterior.\footnote{This is similar in spirit to using point sources for solving the Dirac equation on a lattice-QCD ensemble, where, in principle, the correct limit is already achieved by employing one independent random source position on each configuration.}
This estimator can be improved by using several values of $(r,\theta)$ on each sample, effectively treating the location of $E_0$ within the Wertevorrat on each sample as an additional random variable with a uniform distribution within the Wertevorrat, which is integrated over when evaluating the posterior for $E_0$,
\begin{equation}
    \begin{split}
        \langle E_0 \rangle = \frac{1}{\pi Z} \int d^{n_p}u \int_0^1 dr \, r \int_0^{2\pi} d\theta \, &E_0(\bm{u}, r, \theta) \times {} \\
        &p(\bm{u} | \bm{E}_\mathrm{d}, \bm{u}_0) .
    \end{split}
    \label{eq:WertevorratSampling}
\end{equation}
Here, $E_0(\bm{u}, r, \theta)$ denotes the zero location obtained for the sample parameterized by $\bm{u}$ and using the point at coordinates $(r, \theta)$ within the Wertevorrat as the function estimator.
The uncertainty is estimated similarly, $(\delta E_0)^2 = \langle E_0^2 \rangle - \langle E_0 \rangle^2$.
A schematic representation of the procedure can be found in \cref{fig:sketchWVSampling}.

\begin{figure}
    \centering
    \begin{tikzpicture}
        \tikzset{cross/.style={path picture={
            \draw
            (path picture bounding box.south east) -- (path picture bounding box.north west)
            (path picture bounding box.south west) -- (path picture bounding box.north east);
        }}}

        \node at (1.25,0) {Posterior samples};
        \node at (0,-1) {$\bm u_1$};
        \node at (0,-2) {$\bm u_2$};
        \node at (0,-3) {$\bm u_3$};
        \node[cross] at (0.5,-1.2) {};
        \node[cross] at (1,-1) {};
        \node[cross] at (1.5,-0.7) {};
        \node[cross] at (2,-1.3) {};
        \node[cross] at (2.5,-0.9) {};
        \node[cross] at (0.5,-2.1) {};
        \node[cross] at (1,-1.8) {};
        \node[cross] at (1.5,-2.2) {};
        \node[cross] at (2,-1.9) {};
        \node[cross] at (2.5,-2) {};
        \node[cross] at (0.5,-2.8) {};
        \node[cross] at (1,-3) {};
        \node[cross] at (1.5,-3.2) {};
        \node[cross] at (2,-2.9) {};
        \node[cross] at (2.5,-3.1) {};
        \node at (1.25,-3.75) {\vdots};
        \draw[thick] (-0.2,-1.6) rectangle (2.7,-2.4);

        \draw[thick, ->] (2.7,-2) -- (3.8,-2);
        \node at (6.1,-2) {$\mathcal{K}^{-1}(E) = \mathcal{K}^{-1}_{N,\uline{\textcolor{red}{(r,\theta)}}}(E)$};
        \draw[thick, ->] (6.85,-2.3) -- (6.85,-3);

        \draw[dashed] (6.85,-4.5) circle (1.2);
        \node[circle, fill, inner sep=1pt] at (6.85,-4.5) {};
        \draw[<->] (5.879179606750064,-5.205342302750967) -- (6.85,-4.5) node[midway, below right, inner sep=0pt] {$r_N(\zeta)$};
        \node[cross, red, thick] at (7.12811529493745,-3.644049135334362) {};
        \draw[<->] (7.12811529493745,-3.644049135334362) -- (6.85,-4.5) node[midway, above left, inner sep=0pt] {$\textcolor{red}{r} r_N(\zeta)$};
        \draw[dotted] (6.85,-4.5) -- (7.75,-4.5);
        \draw (7.33,-4.5) arc[start angle=0, end angle=72, radius=0.48];
        \node[above right] at (6.85,-4.5) {$\textcolor{red}{\theta}$};
        \node[left] at (6.85,-4.5) {$c_N(\zeta)$};
        \node[align=right] at (3.95,-4.5) {WV in unit disk\\ coordinates for a\\ specific value of $\zeta$\\ corresponding to $E$};

        \node at (4,-6.2) {$\rightarrow$ find $E_0$ with $\mathcal{M}^{-1}(E_0) = \mathcal{K}^{-1}_{N,\textcolor{red}{(r,\theta)}}(E_0) - i\rho = 0$};
    \end{tikzpicture}
    \caption{Sketch of the uncertainty analysis procedure described in \cref{sec:pole_locations_Nevanlinna_uncertainty}.
             For a fixed sample from the posterior distribution, the function $\mathcal{K}^{-1}$ is defined by a reference point with coordinates $(r, \theta)$ within the Wertevorrat of the Nevanlinna interpolation of $(\bm E, \bm u)$.
             The coordinates $(r, \theta)$ are chosen from a uniform distribution over the unit disk.
             They are held fixed independent of $E$, while the location $c_N$ and size $r_N$ of the Wertevorrat obviously depend on $E$.}
    \label{fig:sketchWVSampling}
\end{figure}

In principle, the uncertainty associated with the analytic continuation via the Wertevorrat may be systematically reduced by using more interpolation nodes.
However, a trade-off exists between reducing uncertainties and facing demands for higher numerical input precision.
In many cases of practical interest, the systematic uncertainty from analytic continuation is smaller than other sources of uncertainty.

\section{Numerical example with toy noisy data}
\label{sec:toy_example}

This section presents an example of the proposed methodology using toy data with Gaussian noise to demonstrate the details of practical implementation and that the method delivers correct results in a controlled setup.

In particular, we consider $p$-wave scattering of two identical particles with mass $m$, so as to correspond roughly to the $\rho$-meson resonance in isospin-1 $p$-wave $\pi\pi$ scattering on the $m_\pi = 283~\mathrm{MeV}$ HadSpec ensembles~\cite{Rodas:2023gma}. The phase shift is defined by a Breit-Wigner curve,
\begin{equation}
    \left( \frac{q}{m} \right)^3 \cot\delta_1(q^2) = 6\pi E \frac{m_\mathrm{BW}^2 - E^2}{g^2} ,
    \label{eq:phaseshiftBW}
\end{equation}
with Breit-Wigner mass $m_\mathrm{BW}$ and coupling $g$. We choose $m_\mathrm{BW} / m = 0.13440/0.04720 \approx 2.85$ and $g = 5.56$, corresponding to the best Breit-Wigner fit to the data in Ref.~\cite{Rodas:2023gma}.
The associated scattering amplitude has a resonance pole at $E_\mathrm{pole, true}/m \approx 2.8308 - i0.1023$.

We solve the corresponding quantization condition \cite{Dudek:2012xn,Morningstar:2017spu} for the finite-volume energy levels of this system in two volumes $mL = 3.9$ and 5.2, as well as for 5 distinct frames, \ie those with $\bm P^2 < 5 (2\pi/L)^2$.
In each frame, we only use a single irrep, $T_1^-$ in the rest frame and $A_1$ in each of the moving frames.
We only keep levels with $E/m < 4$.
This procedure results in a total of 24 energy levels.
We then add uncorrelated random Gaussian noise to these data points.

\begin{figure}
    \centering
    \includegraphics[width=1\linewidth]{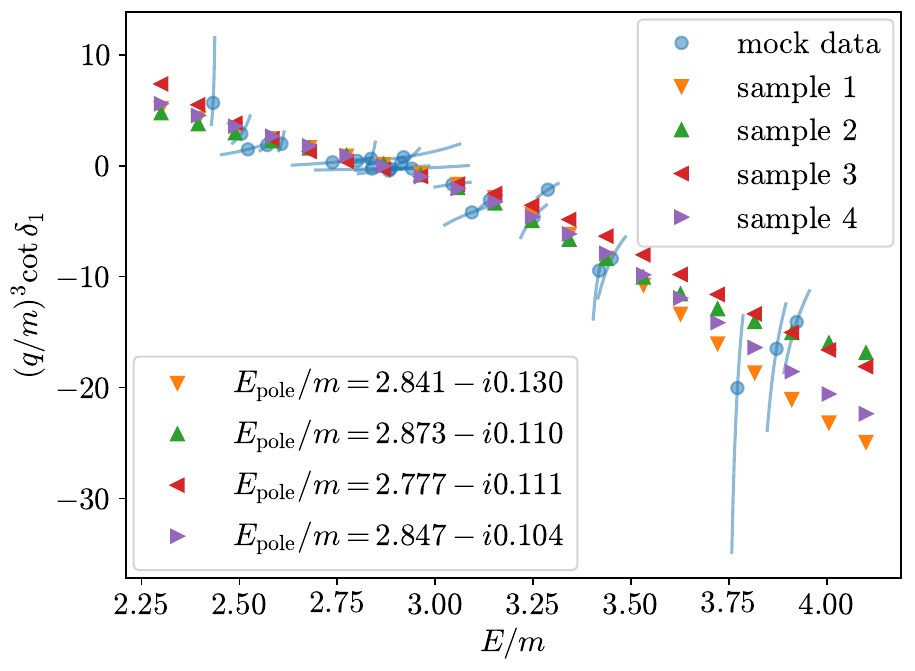}
    \caption{Numerical example based on the toy resonance presented in \cref{sec:toy_example}.
      The blue points in the background show the noisy mock data generated as detailed in the text.
      Each set of triangles represents an independent sample drawn from the (Gaussian approximation of the) posterior distribution.
      For each sample, the pole position is added as a reference.}
    \label{fig:phaseshiftSamplesBW}
\end{figure}

\begin{figure*}
    \centering
    \includegraphics[width=0.7\linewidth]{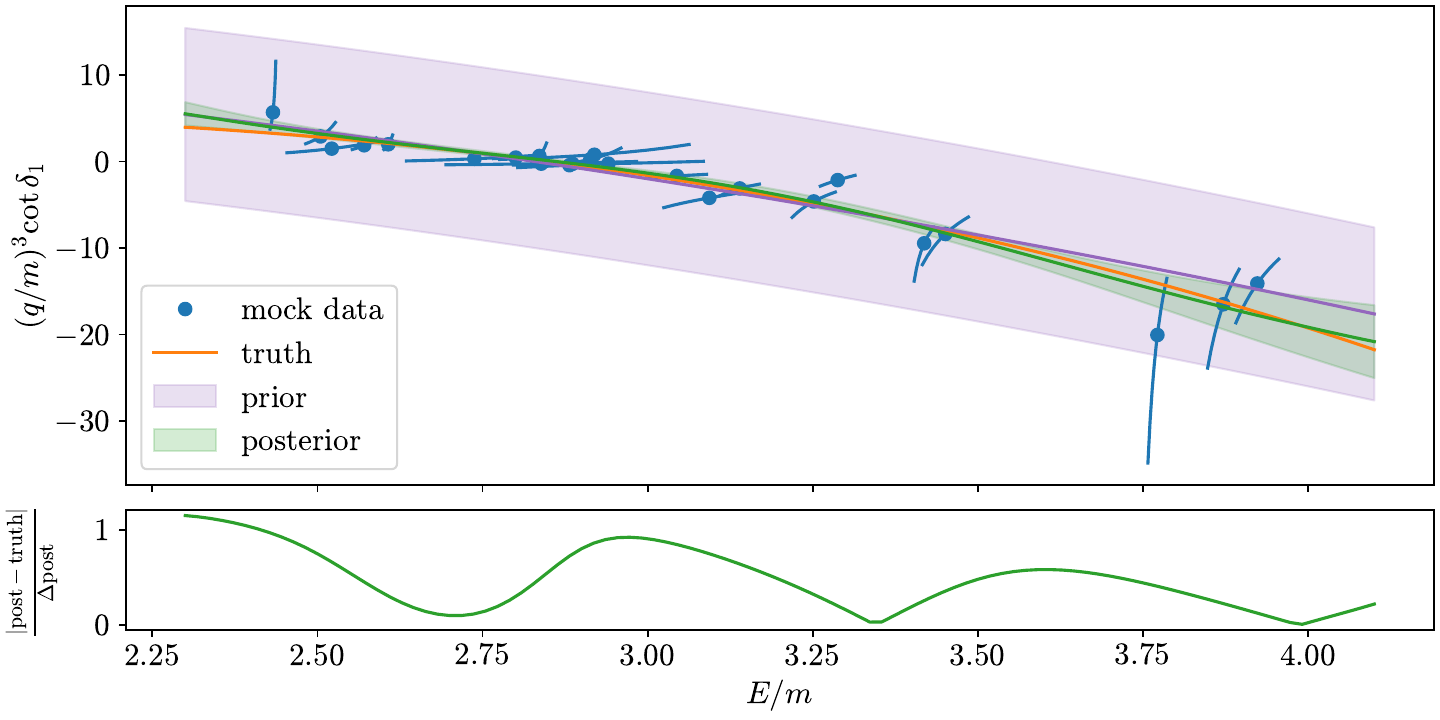}
    \caption{Phase shift for the model in \cref{eq:phaseshiftBW} (orange line).
      The blue points show the noisy mock data generated as detailed in the text, while the purple line and band depict the prior.
      The green line and band indicate the posterior central value according to \cref{eq:expectedu} and the Bayes uncertainty according to \cref{eq:erroru}. The Monte-Carlo integration error as well as the size of the Wertevorrat are negligibly small.
      Note that the effect of $\ell_c$ is not visible in this graphical representation of the prior (purple), although it does affect the results of the Bayesian analysis, \ie the posterior (green).
      The lower panel displays the absolute difference between the posterior and the truth in units of the Bayes error.
      }
    \label{fig:phaseshiftBW}
\end{figure*}

For the Bayesian analysis, we discretize the energy as 20 equidistant nodes with $E/m \in [2.3, 4.1]$. In the case of $p$-wave scattering considered in this section, it is standard to parameterize $u(E) =  (q/m)^3 \cot\delta_1$. As a prior, we choose a simple phase-shift function with a linear dependence on $q^2$,
\begin{equation}
    u_0(E) = A + B \left(\frac{q}{m}\right)^2 = A + B \left[ \frac{E^2}{4m^2} - 1 \right] ,
    \label{eq:priorLinear}
\end{equation}
with $A = -B = 8$.
This prior roughly follows the data.
The covariance of the prior is parameterized using an RBF kernel according to \cref{eq:pprior} with $\sigma = 10$, $\ell_c/m = 0.7$, and a regulator of $\epsilon = 10^{-6}$.
As can be seen from \cref{fig:phaseshiftSamplesBW,fig:phaseshiftBW}, the maximum distance in energy between any two data points is $\Delta E_\mathrm{max}/m \approx 1.08$, so that the value of $\ell_c/m = 0.7$ is of the same order of magnitude and ensures that the function is constrained in these gaps.
In \cref{fig:phaseshiftBW}, it can also be seen that the choice of $\sigma = 10$ is larger than the typical variation of the data points along the vertical axis.

In order to sample the posterior distribution, we employ the reweighting procedure outlined in \cref{sec:posteriorSampling}.
The expected sample size turns out to be $\mathrm{ESS} = \langle w_* \rangle_*^2 / \langle w_*^2 \rangle_* \approx 68\%$, which indicates that the approximate distribution $p^*$ (\cf \cref{eq:approximateProbability}) is sufficiently close to the true posterior distribution for the reweighting approach to work well.
Four illustrative samples drawn from the approximate distribution are shown together with the mock data in \cref{fig:phaseshiftSamplesBW}.
This illustrates that our non-parametric Bayesian approach allows us to sample a broad range of functional behaviors, which is an inherent difference to performing fits to the data, where each fit is restricted to a specific functional form.
The pole positions corresponding to the shown samples are also indicated in \cref{fig:phaseshiftSamplesBW}.

The expected posterior and its uncertainty are then determined by averaging over the reweighted samples, following \cref{eq:expectedu,eq:erroru}.
The resulting posterior is plotted together with the mock data, the model truth, and the prior in the upper panel of \cref{fig:phaseshiftBW}.
It is clearly visible that the posterior follows the form of \cref{eq:phaseshiftBW} closely in the energy interval considered.
The difference between the posterior and the model truth is typically well within $1\,\sigma$, as shown in the lower panel of \cref{fig:phaseshiftBW}.
We also monitored the total data $\chi^2$ of the posterior samples and found $\chi^2 \sim n_\mathrm{d}$, indicating reasonable fits.

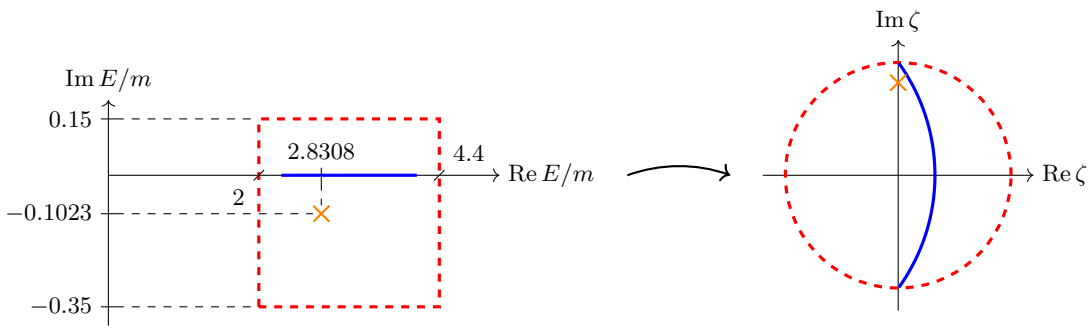
\begin{figure*}
    \centering
    \begin{tikzpicture}
        \tikzset{cross/.style={path picture={
            \draw
            (path picture bounding box.south east) -- (path picture bounding box.north west)
            (path picture bounding box.south west) -- (path picture bounding box.north east);
        }}}

        \draw[->] (0, -2) -- (0, 1) node[above] {$\imag E/m$};
        \draw[->] (0, 0) -- (5.2, 0) node[right] {$\real E/m$};
        \draw[very thick, blue] (2.3, 0) -- (4.1, 0);
        \draw[very thick, red, dashed] (2, -1.75) rectangle (4.4, 0.75);
        \draw[-] (-0.1, -1.75) node[left] {$-0.35$} -- (0.1, -1.75);
        \draw[-] (-0.1, 0.75) node[left] {$0.15$} -- (0.1, 0.75);
        \draw[-] (-0.1, -0.5115) node[left] {$-0.1023$} -- (0.1, -0.5115);
        \draw[-] (1.93, -0.07) node[below left] {$2$} -- (2.07, 0.07);
        \draw[-] (4.33, -0.07) -- (4.47, 0.07) node[above right] {$4.4$};
        \draw[-] (2.8308, -0.1) -- (2.8308, 0.1) node[above] {$2.8308$};
        \node[cross, thick, orange] at (2.8308, -0.5115) {};
        \draw[thin, dashed] (0.1, -1.75) -- (2, -1.75);
        \draw[thin, dashed] (0.1, 0.75) -- (2, 0.75);
        \draw[thin, dashed] (0.1, -0.5115) -- (2.72, -0.5115);
        \draw[thin, dashed] (2.8308, -0.1) -- (2.8308, -0.4);

        \draw[thick, ->] (6.9, 0) arc (110:70:2);

        \draw[->] (8.7, 0) -- (12.3, 0) node[right] {$\real \zeta$};
        \draw[->] (10.5, -1.8) -- (10.5, 1.8) node[above] {$\imag \zeta$};
        \draw[very thick, red, dashed] (10.5, 0) circle (1.5);
        \draw[very thick, blue] (10.509451, 1.491693) -- (10.510172, 1.490645) -- (10.510982, 1.489476) -- (10.511891, 1.488170) -- (10.512912, 1.486710) -- (10.514057, 1.485078) -- (10.515339, 1.483252) -- (10.516774, 1.481210) -- (10.518380, 1.478924) -- (10.520177, 1.476366) -- (10.522185, 1.473503) -- (10.524428, 1.470298) -- (10.526932, 1.466711) -- (10.529726, 1.462696) -- (10.532841, 1.458204) -- (10.536311, 1.453176) -- (10.540173, 1.447552) -- (10.544469, 1.441261) -- (10.549241, 1.434226) -- (10.554537, 1.426361) -- (10.560407, 1.417571) -- (10.566904, 1.407752) -- (10.574082, 1.396786) -- (10.582001, 1.384548) -- (10.590719, 1.370896) -- (10.600294, 1.355678) -- (10.610786, 1.338728) -- (10.622249, 1.319864) -- (10.634735, 1.298892) -- (10.648285, 1.275604) -- (10.662933, 1.249776) -- (10.678696, 1.221177) -- (10.695574, 1.189562) -- (10.713541, 1.154684) -- (10.732547, 1.116292) -- (10.752504, 1.074140) -- (10.773288, 1.027995) -- (10.794733, 0.977641) -- (10.816623, 0.922900) -- (10.838698, 0.863631) -- (10.860647, 0.799755) -- (10.882118, 0.731262) -- (10.902719, 0.658226) -- (10.922033, 0.580817) -- (10.939630, 0.499308) -- (10.955087, 0.414084) -- (10.968006, 0.325636) -- (10.978035, 0.234559) -- (10.984889, 0.141535) -- (10.988368, 0.047313) -- (10.988368, -0.047313) -- (10.984889, -0.141535) -- (10.978035, -0.234559) -- (10.968006, -0.325636) -- (10.955087, -0.414084) -- (10.939630, -0.499308) -- (10.922033, -0.580817) -- (10.902719, -0.658226) -- (10.882118, -0.731262) -- (10.860647, -0.799755) -- (10.838698, -0.863631) -- (10.816623, -0.922900) -- (10.794733, -0.977641) -- (10.773288, -1.027995) -- (10.752504, -1.074140) -- (10.732547, -1.116292) -- (10.713541, -1.154684) -- (10.695574, -1.189562) -- (10.678696, -1.221177) -- (10.662933, -1.249776) -- (10.648285, -1.275604) -- (10.634735, -1.298892) -- (10.622249, -1.319864) -- (10.610786, -1.338728) -- (10.600294, -1.355678) -- (10.590719, -1.370896) -- (10.582001, -1.384548) -- (10.574082, -1.396786) -- (10.566904, -1.407752) -- (10.560407, -1.417571) -- (10.554537, -1.426361) -- (10.549241, -1.434226) -- (10.544469, -1.441261) -- (10.540173, -1.447552) -- (10.536311, -1.453176) -- (10.532841, -1.458204) -- (10.529726, -1.462696) -- (10.526932, -1.466711) -- (10.524428, -1.470298) -- (10.522185, -1.473503) -- (10.520177, -1.476366) -- (10.518380, -1.478924) -- (10.516774, -1.481210) -- (10.515339, -1.483252) -- (10.514057, -1.485078) -- (10.512912, -1.486710) -- (10.511891, -1.488170) -- (10.510982, -1.489476) -- (10.510172, -1.490645) -- (10.509451, -1.491693);
        \node[cross, thick, orange] at (10.499136317820758, 1.2315311807774936) {};
    \end{tikzpicture}
    \caption{Mapping of the complex energy plane (left panel) to the unit disk (right panel) for the Breit-Wigner toy example.
        The blue line indicates the interval of energies which we discretize for the Bayesian analysis.
        The red rectangle shows the patch of the complex plane which we map to the unit disk.
        The true pole position is marked by the orange cross. Note that the blue line on the right panel does not intersect the boundary of the disk.}
    \label{fig:mappingBW}
\end{figure*}

Next we extract the pole position from this toy-data example using Nevanlinna--Pick interpolation as explained in \cref{sec:analytic_continuation}.
For this purpose, we map the complex energy plane (in mass units) to the unit disk, employing the Schwarz--Christoffel rectangle mapping (\cf \cref{sec:schwarz_christoffel}).
We choose a rectangle centered at $C = 3.2 - 0.1i$, and with a total width of $W = 2.4$ and a full height of $H = 0.5$.
Thus, along the real axis, our mapping region is centered with our considered energy interval, while along the imaginary axis, it is slightly shifted towards the expected pole location.
This rectangle is marked in red in a sketch of the complex energy plane in \cref{fig:mappingBW}.
Here, we also indicate the true pole position (orange cross), which lies well within the rectangle.
The size of the codomain to be mapped to the unit disk is determined as described in \cref{sec:mappings_general}.

The pole locations and their uncertainties are determined as described in \cref{sec:pole_locations_Nevanlinna_uncertainty}, using one sample from the Wertevorrat on each configuration.
We constrain our algorithm to only look for poles on the unphysical sheet with $\real q \geq 0$ and $\imag q \leq 0$, and such that the corresponding energy lies within the rectangle which is mapped to the unit disk.
In particular, we use the Trust Region Reflective method \cite{Branch1999} to minimize $|(q/m) \cot \delta_1(q) - iq|$ subject to the aforementioned bounds and check that the value at the solution is indeed zero within tolerance.
We try different initial guesses surrounding the expected pole location until we find a solution which satisfies the above criteria.

We find a resonance pole on each of our samples from the posterior.
For the expectation value of the pole position, this procedure yields
\begin{equation}
    E_\mathrm{pole}/m = 2.833(4)(40) - i0.119(3)(29).
\end{equation}
Here, the first error refers to the Monte-Carlo integration error from sampling the posterior distribution, and the second one to the combined Bayes and Wertevorrat uncertainties.
The Bayes error includes both the statistical error from the Monte-Carlo sampling of the gauge field and the parameterization dependence of the K matrix.
The Wertevorrat, on the other hand, encodes the uncertainty from the analytic continuation.
The latter two are combined into one error by means of the sampling approach explained in \cref{sec:pole_locations_Nevanlinna_uncertainty}.
The real and imaginary parts of this result both agree with the true pole position quoted above within $1~\sigma$.

\begin{figure}
    \centering
    \includegraphics[width=1\linewidth]{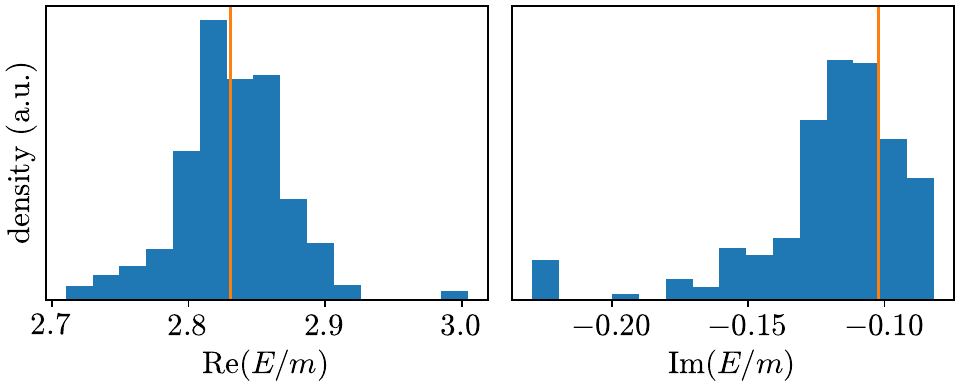}
    \caption{Histograms of the real (left) and imaginary (right) parts of the pole location for the toy noisy data in \cref{sec:toy_example}.
        The orange lines show the true pole location of the scattering amplitude defined by the phase shift in \cref{eq:phaseshiftBW}.
        The vertical axis is expressed in arbitrary units (a.u.).
    }
    \label{fig:polePositionsHistogramsBW}
\end{figure}

In order to visualize the distribution of pole locations we obtain from our analysis of the toy noisy data, we show histograms of the real and imaginary parts of the pole location in \cref{fig:polePositionsHistogramsBW}.
Plotted are the histograms over the configurations drawn from the posterior distribution, taking the reweighting which is necessary in our sampling approach properly into account (\cf \cref{sec:posteriorSampling}).
The true pole location is marked in orange in \cref{fig:polePositionsHistogramsBW}.
This illustrates again the good agreement between the posterior distribution and the known answer.

\section{Applications to the analysis of lattice QCD spectra}
\label{sec:analysisqcd}

After having discussed a toy example in the previous section, we now turn to the application of our method to actual lattice-QCD data.
In the following, we will consider three different systems: $\pi\pi$ scattering at maximal isospin ($I = 2$) on the CLS ensemble D200, $\pi\pi$ scattering at isospin $I = 0$ on HadSpec ensembles, and coupled-channel $\pi\Sigma - \bar{K}N$ scattering on D200.
The first example does not feature any interesting physics in the complex plane, but serves as an illustrative example of the Bayesian analysis on the real line.
While the first two cases can be treated as single-channel systems, the last example showcases the straightforward generalization of our method to coupled-channel scattering.

All ensembles considered in the present paper have $2 + 1$ flavors of dynamical quarks.
The CLS ensemble D200 has a pion mass around 200~MeV and a kaon mass around 480~MeV, at a lattice spacing of $a \approx 0.06~\mathrm{fm}$ and a physical volume of about $(4~\mathrm{fm})^3$ \cite{Bruno:2014jqa,Blanton:2021llb,Dawid:2025doq}.
From HadSpec, we analyze two anisotropic lattices, both at a pion mass of 283~MeV and a kaon mass of 519~MeV, with a spatial lattice spacing of $a_s \approx 0.12~\mathrm{fm}$ and an anisotropy of $\xi \approx 3.457$.
The two HadSpec ensembles we consider have physical volumes of about $(2.88~\mathrm{fm})^3$ and $(3.84~\mathrm{fm})^3$, respectively \cite{Edwards:2008ja,HadronSpectrum:2008xlg,Rodas:2023gma}.
For further ensemble details, we refer to the original papers.

\subsection{Maximal-isospin \texorpdfstring{$\pi\pi$}{pi pi} systems on D200}
\label{sec:2piI2}
The scattering of two and three mesons at maximal isospin has been studied on several CLS ensembles in Refs.~\cite{Blanton:2019vdk,Blanton:2021llb,Draper:2023boj,Dawid:2025doq,Dawid:2025zxc}.
In the following, we reanalyze their data for the $s$-wave scattering of two pions at the lightest quark mass considered in Ref.~\cite{Blanton:2021llb}, \ie on the ensemble D200 with $m_\pi \approx 200~\mathrm{MeV}$.
Since the scattering amplitude of this system does not have any poles in the relevant energy region, our objective here is to reproduce their results for the scattering length and effective range, giving an improved estimate of the systematic uncertainty. While Ref.~\cite{Blanton:2021llb} considered $s$ and $d$ waves, the effect of $d$ waves was found to be small on the D200 ensemble, and as such, we perform an $s$-wave only analysis.

\begin{figure*}
    \centering
    \subfloat{\includegraphics[width=0.5\linewidth]{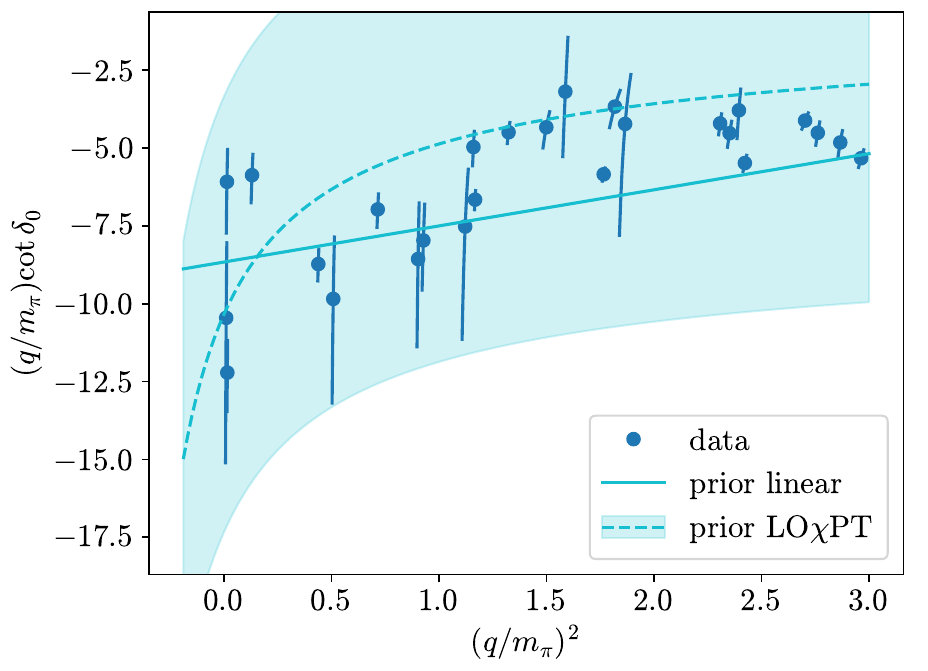}}
    \subfloat{\includegraphics[width=0.5\linewidth]{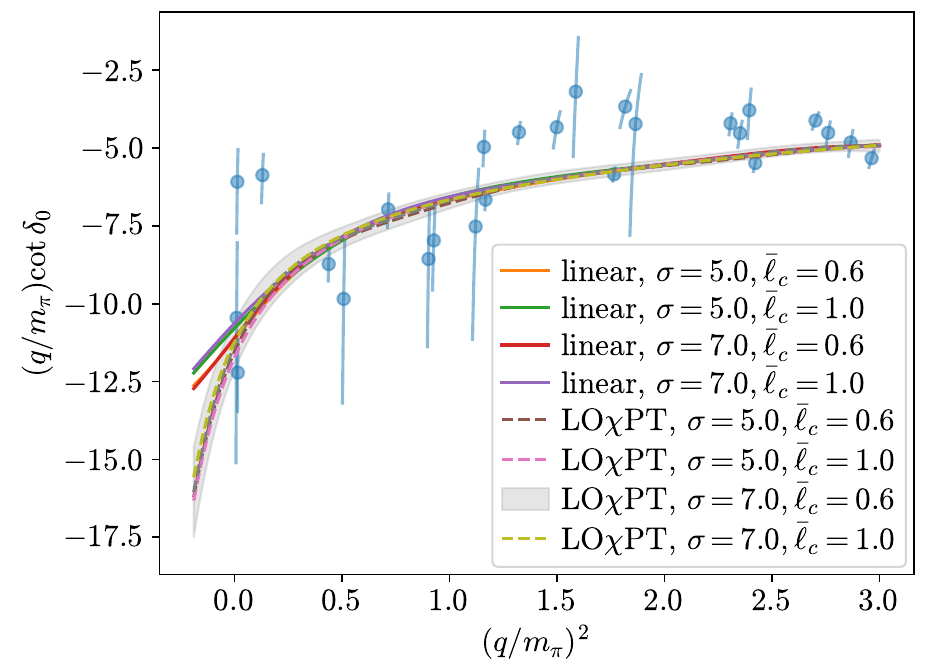}}
    \caption{Priors (left panel) and posteriors (right panel) for the phase shift of the isospin-2 $\pi\pi$ system.
      We compare the two classes of priors defined by \cref{eq:priorLinear} (solid lines) and \cref{eq:prior2piI2LOChPT} (dashed lines), respectively, and different choices for the RBF kernel hyperparameters $\sigma$ and $\bar{\ell}_c = \ell_c/m_\pi$.
      For visual clarity, only one of the priors and the corresponding posterior are plotted with an error band.
      The Bayesian error bands of the other posteriors on the right panel are of comparable width.
      }
    \label{fig:phaseshift2piI2}
\end{figure*}

The first step in our Bayesian analysis again consists in the discretization of the energy.
Here, we use a grid of 21 equidistant nodes with $E/m_\pi \in [1.8, 4]$.
We consider two different classes of priors: a simple prior linear in $q^2$ according to \cref{eq:priorLinear},
with $A = -8.66$ and $B = 1.16$, and a prior given by the leading-order (LO) $\chi$PT expression for $(q/m_\pi) \cot\delta_0$ \cite{Weinberg:1966kf,Gasser:1983yg},
\begin{equation}
    u_0(E) = -16\pi \left(\frac{F_\pi}{m_\pi}\right)^2 \frac{E m_\pi}{E^2 - 2m_\pi^2} .
    \label{eq:prior2piI2LOChPT}
\end{equation}
Here, we set $F_\pi = 92.2~\mathrm{MeV}$, which corresponds in pion-mass units on D200 to roughly $F_\pi / m_\pi \approx 0.453$ \cite{Ce:2022eix}.
The form \cref{eq:prior2piI2LOChPT} features an Adler zero, \ie a zero of the scattering amplitude below threshold, at $(E/m_\pi)^2 = 2$ \cite{Adler:1964um}.
The two priors are shown together with the original data in \cref{fig:phaseshift2piI2} (left).
As done in Refs.~\cite{Blanton:2021llb,Draper:2023boj,Dawid:2025doq}, we use the covariance of the energy shifts with respect to the non-interacting levels (rather than the covariance of the absolute energies) as $\Sigma_\mathrm{d}$.
This amounts to neglecting the uncertainties of the single-hadron energies. They can be factored in at a later stage by understanding the results for the pole locations from this procedure to be relative to the lowest relevant threshold. The full errors of the absolute pole positions can be approximated by adding the threshold error in quadrature.

In \cref{fig:phaseshift2piI2} (right), we compare the posteriors obtained with these two priors as well as with different RBF kernels ($\sigma = 5$ or 7, and $\ell_c/m_\pi = 0.6$ or 1.0).
These values of $\sigma$ are chosen to be significantly larger than the spread of the data, while the choice of $\ell_c$ is determined by the density in energy of the lattice-QCD energy levels.

As can be seen from \cref{fig:phaseshift2piI2} (right), many of the data points differ by more than $1\,\sigma$ from the posterior. Accordingly, we find rather large values for the data $\chi^2$, with $\chi^2 \sim 2.5n_\mathrm{d}$. This is in qualitative agreement with the reduced $\chi^2$ values found for parametric fits on this ensemble in Ref.~\cite{Blanton:2021llb}.

While all the posteriors are very similar for $(q/m_\pi)^2 \gtrsim 0.1$, at lower energies the ones obtained from the LO$\chi$PT priors start to fall off more steeply than the ones obtained from the linear priors.
The value at threshold, \ie the scattering length, is still in good agreement; the derivative, \ie the effective range, however, differs already significantly here.

This can be seen even more clearly from \cref{tab:2piI2results}, where we list the scattering length and the product of scattering length and effective range (\cf \cref{eq:ERE}) for the different priors from \cref{fig:phaseshift2piI2}.
Since we only know the posterior at discrete values of the energy, it is clear that these observables cannot be calculated directly from our result, but require an interpolation.
We again use Nevanlinna--Pick interpolation combined with a Schwarz--Christoffel rectangle mapping.
In energy coordinates, we choose a rectangle centered with our considered energy interval, $C = 2.9$, and with a total width of $W = 2.8$ and a total height of $H = 0.6$.
The size of the codomain to be mapped to the unit disk is again determined as described in \cref{sec:mappings_general}.
The scattering length is then straightforwardly obtained by evaluating the interpolating function at $E/m_\pi = 2$, while for the calculation of the effective range, we use a central finite-difference approximation to the derivative.
For both observables, the uncertainty due to the interpolation (\ie the Wertevorrat) can be incorporated analogously to the case of determining pole positions, \ie by sampling from the Wertevorrat as explained in \cref{sec:pole_locations_Nevanlinna_uncertainty}.
However, since both $a_0$ and $r_0$ are defined for real energies, this is a pure interpolation problem, and consequently, the Wertevorrat uncertainty is negligible.

From \cref{tab:2piI2results}, it can be seen that the results from the analyses using either the linear or the $\chi$PT prior agree better the larger $\sigma$ and the smaller $\ell_c$ is, \ie the fewer assumptions one introduces with the prior.
The errors also tend to increase in these directions, which highlights the importance of carefully comparing different choices for the prior covariance.
We remark that the Bayes factors are only meaningful for comparing different prior central values at fixed $\sigma$ and $\ell_c$, as discussed in \cref{sec:prior_dependence}.
As can be seen from \cref{tab:2piI2results}, at fixed prior covariance matrix, the $\chi$PT prior is always favored over the linear one.
This means that the $\chi$PT function is generally favored by the data, as has been observed as well in the fitting approach \cite{Blanton:2019vdk,Fischer:2020jzp,Blanton:2021llb}.

\begin{table}[htb]
    \caption{Results for the isospin-2 $\pi\pi$ system with different priors.
        The Bayes factors are normalized for each combination of $(\sigma, \ell_c)$ separately with respect to the larger one of the two priors considered here.
        The first error refers to the Monte-Carlo integration error and the second one to the combined Bayes and Wertevorrat uncertainties.
        The preferred analysis is highlighted in gray.
        }
    \label{tab:2piI2results}
    \begin{ruledtabular}
        \begin{tabular}{lcccll}
            Prior      & $\sigma$ & $\ell_c/m_\pi$ & Bayes factor & \multicolumn{1}{c}{$m_\pi a_0$} & \multicolumn{1}{c}{$m_\pi^2 a_0 r_0$} \\ \hline
            Linear     & 5.0      & 0.6            & 0.558        & $0.0915(4)(65)$                 & $1.37(3)(52)$ \\
            LO$\chi$PT & 5.0      & 0.6            & 1.000        & $0.0886(4)(64)$                 & $2.54(3)(48)$ \\[\defaultaddspace]

            Linear     & 5.0      & 1.0            & 0.213        & $0.0947(5)(66)$                 & $1.27(2)(28)$ \\
            LO$\chi$PT & 5.0      & 1.0            & 1.000        & $0.0880(6)(66)$                 & $2.52(2)(24)$ \\[\defaultaddspace]

            Linear     & 7.0      & 0.6            & 0.638        & $0.0904(6)(67)$                 & $1.56(5)(66)$ \\
            \rowcolor{gray!25}[2\tabcolsep]
            LO$\chi$PT & 7.0      & 0.6            & 1.000        & $0.0882(6)(66)$                 & $2.65(5)(62)$ \\[\defaultaddspace]

            Linear     & 7.0      & 1.0            & 0.348        & $0.0934(4)(62)$                 & $1.40(2)(34)$ \\
            LO$\chi$PT & 7.0      & 1.0            & 1.000        & $0.0884(4)(62)$                 & $2.51(2)(30)$
        \end{tabular}
    \end{ruledtabular}
\end{table}

Ref.~\cite{Dawid:2025doq} performed an AIC average over a range of fits, yielding $m_\pi a_0 = 0.0886(57)$ and $m_\pi^2 a_0 r_0 = 2.42(34)$.
Comparing to these numbers, one also notices that our results using the $\chi$PT prior are in better agreement with the fit-based result than are those using the linear prior.
We remark that all fits considered in Refs.~\cite{Blanton:2021llb,Dawid:2025doq} are based on the $\chi$PT form,\footnote{Refs.~\cite{Blanton:2021llb,Dawid:2025doq} make the prefactor (and possibly the location of the Adler zero) a free parameter, and enhance the formula by additional $q^2$-dependent terms.} \ie they enforce explicitly an Adler zero.
Moreover, the majority of their fits include three-meson data that we do not analyze here. Hence, one would not expect to get exactly the same answer from our analysis, even though it should be statistically compatible. The increase in error which we observe in comparison to Ref.~\cite{Dawid:2025doq} is thus most likely also due to the smaller dataset.

As final results for the $s$-wave analysis of the $\pi^+\pi^+$ data, we choose those from the LO$\chi$PT prior with the largest $\sigma$ and smallest $\ell_c$ from \cref{tab:2piI2results}, \ie
\begin{align}
    m_\pi a_0 &= 0.0882(6)(66), \\
    m_\pi^2 a_0 r_0 &= 2.65(5)(62).
\end{align}

\subsection{Isospin-0 \texorpdfstring{$\pi\pi$}{pi pi} system on \texorpdfstring{$m_\pi = 283~\mathrm{MeV}$}{Mpi = 283 MeV} ensembles by HadSpec}

The HadSpec Collaboration has studied $\pi\pi$ scattering at all isospins in Ref.~\cite{Rodas:2023gma}.
Here, we are particularly interested in the isospin-0 channel, where the broad $\sigma$ meson resonance is present for physical quark masses, the pole position of which is notoriously difficult to pin down \cite{Caprini:2005zr,Hanhart:2008mx}.
In the following, we will reanalyze the energy-level data from Ref.~\cite{Rodas:2023gma} on the two ensembles with $m_\pi = 283~\mathrm{MeV}$ and spatial extents of $L\approx 2.88~{\rm fm}$ and $L\approx 3.84~{\rm fm}$.
We will compare the results of our approach to those in Ref.~\cite{Rodas:2023gma} based on fits to parametric models for the K matrix.

The analysis begins with a discretization of the real-energy axis using 24 nodes with $E/m_\pi \in [1.6, 3.5]$, roughly corresponding to the location of the finite-volume energy levels.
To obtain the tightest constraints near the expected region of the $\sigma$ pole, a node spacing of 0.05 is used for $E/m_\pi \in [1.6, 2]$, while a wider spacing of 0.1 is used for $E/m_\pi \in [2, 3.5]$.

\begin{figure*}
    \centering
    \subfloat{\includegraphics[width=0.5\linewidth]{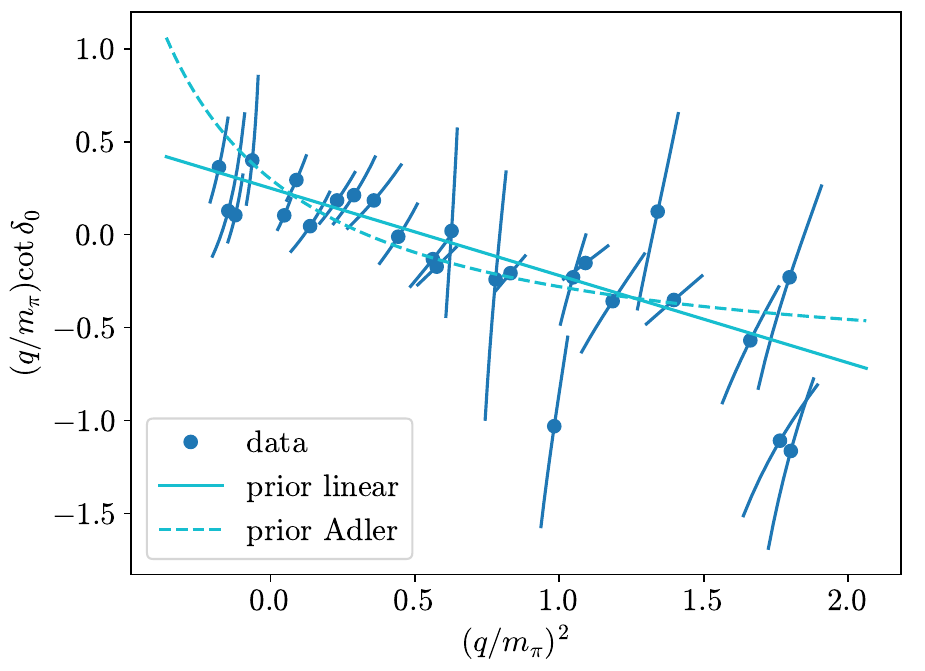}}
    \subfloat{\includegraphics[width=0.5\linewidth]{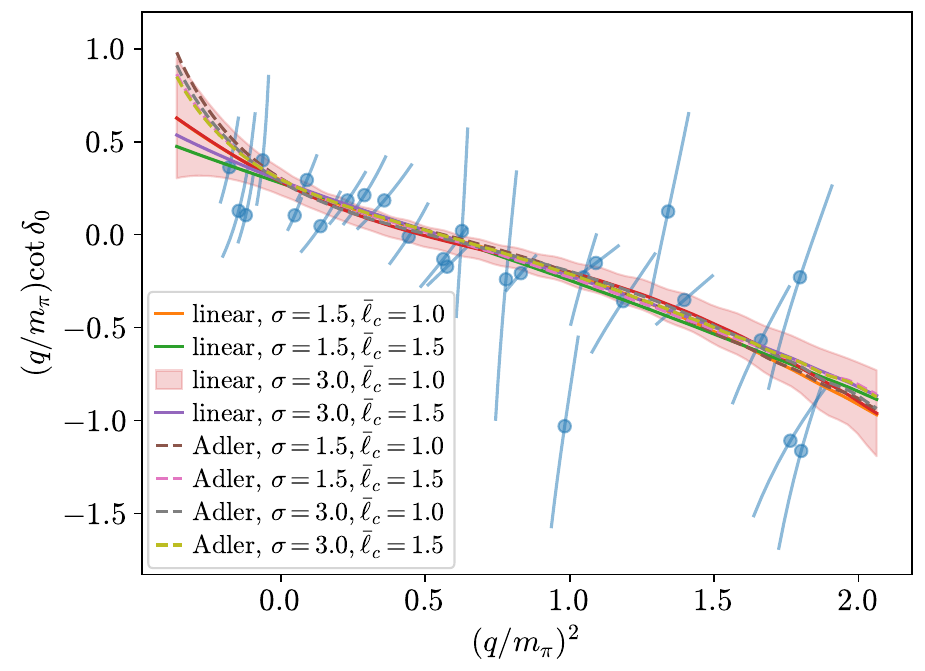}}
    \caption{Priors (left panel) and posteriors (right panel) for the phase shift of the isospin-0 $\pi\pi$ system.
      We compare the two classes of priors defined by \cref{eq:priorLinear} (solid lines) and \cref{eq:priorSigmaAdler} (dashed lines), respectively, and different choices for the RBF kernel hyperparameters $\sigma$ and $\bar{\ell}_c = \ell_c/m_\pi$.
      For visual clarity, only one of the posteriors is plotted with an error band on the right panel. The error bands of the other posteriors are of comparable width.
      Note that on the left panel, the width of the priors exceeds the range shown in the plot.
      }
    \label{fig:phaseshiftSigma}
\end{figure*}

Again we consider two different classes of priors: a simple prior linear in $q^2$ according to \cref{eq:priorLinear} with $A = 0.25$ and $B = -0.47$, and a prior explicitly featuring an Adler zero \cite{Adler:1964um}, \ie a subthreshold zero in the scattering amplitude,
\begin{equation}
    u_0(E) = \frac{4m_\pi^2 - s_A}{E^2 - s_A} \left[ A + B\left(\frac{q}{m_\pi}\right)^2 \right] ,
    \label{eq:priorSigmaAdler}
\end{equation}
where the location of the Adler zero is taken from the leading-order prediction of $\chi$PT,
$s_A = m_\pi^2/2$.
The values for the other free parameters ($A = 0.3$, $B = -0.9$) have been chosen in both cases so that the priors roughly encompass the phase-shift values predicted by the data.
The two priors are shown together with the original data from Ref.~\cite{Rodas:2023gma} in the left panel of \cref{fig:phaseshiftSigma}.
We will compare results for the two choices as well as for different RBF kernels below.

In order to map the complex energy plane (in pion-mass units) to the unit disk, we again employ the Schwarz--Christoffel rectangle mapping.
Our default rectangle is centered at $C = 2.55 - 0.3i$, has a total width of $W = 2.6$, and a total height of $H = 0.7$.
Depending on the choice of prior, this mapping yields a solution for a zero in the inverse scattering amplitude (\ie, for the $\sigma$ pole position) on $97\%-100\%$ of samples.
On the few remaining samples, we try mapping domains shifted either farther into the complex plane, up to $C = 2.55 - 0.5i$ with corresponding $H = 1.1$, or centered around the real axis, \ie with $C = 2.55$ and corresponding $H = 1.3$.
This approach enables the identification of a pole location within the mapping region for each sample from the posterior.
In all cases, the size of the codomain to be mapped to the unit disk is determined as described in \cref{sec:mappings_general}.

The right panel of \cref{fig:phaseshiftSigma} displays the original data from Ref.~\cite{Rodas:2023gma} together with the posterior for the phase shift on the real line, including results from the two choices of prior central values in \cref{eq:priorLinear,eq:priorSigmaAdler} (\cf left panel of \cref{fig:phaseshiftSigma}) as well as different RBF kernels with $\sigma = 1.5$ or 3.0, and with $\ell_c/m_\pi = 1.0$ or 1.5.
Above threshold, results from different priors agree closely, which matches the observations in Ref.~\cite{Rodas:2023gma} regarding the comparison of different fit models.
This agreement indicates that the data constrain the posterior tightly above threshold.
Below threshold, where there are almost no data, the posteriors follow the priors and begin to deviate from each other.
As expected, the posteriors start converging to the priors,
on a length scale governed by the correlation length $\ell_c$,
after the leftmost data point.
However, for the relatively wide priors considered here, the posteriors obtained from the two different prior central values still agree within their Bayes uncertainties, as can also be seen from \cref{fig:phaseshiftSigma}.
Significantly smaller prior widths can lead to an enhanced prior dependence and eventually render the results from different priors incompatible.
Detailed exploration of this regime exceeds the scope of the present work.

\Cref{tab:sigmaResults} gives the probability of finding a virtual bound state versus a resonance as well as the locations of the $\sigma$ pole from different priors.
Pole locations and uncertainties are determined as in \cref{sec:pole_locations_Nevanlinna_uncertainty} using one sample from the Wertevorrat on each configuration.
The integration error (first number in brackets) is always much smaller than the combined Bayesian and Wertevorrat uncertainty (second number in brackets), despite the fact that only 100 posterior samples were used.
Within the quoted uncertainties, a consistent picture of the pole emerges from the different priors.
However, the linear prior seems to slightly favor the virtual-bound-state scenario, while the Adler prior prefers a resonance.
With increasing prior width and decreasing prior correlation length, \ie reducing the strength of the assumptions, the degree of preference for either scenario decreases for both priors, so that they become closer, as already observed in \cref{fig:phaseshiftSigma} on the real line.
The Bayes factors, which are also listed in \cref{tab:sigmaResults}, always favor the linear prior, given a fixed prior covariance matrix.
This is consistent with the findings of Ref.~\cite{Rodas:2023nec}, which indicated that, at this pion mass, the central values of their favored dispersive results do not produce an Adler zero.

\begin{table*}[htb]
    \caption{Results for the isospin-0 $\pi\pi$ system with different priors.
        The Bayes factors are normalized for each combination of $(\sigma, \ell_c)$ separately with respect to the larger one of the two priors considered here.
        For the $\sigma$ pole positions, we quote the probability of finding a virtual bound state (v.b.s.)\ versus a resonance (res.), the overall average of the pole location in the energy plane in pion-mass units, as well as the pole locations for the two cases separately.
        The first error refers to the Monte-Carlo integration error and the second one to the combined Bayes and Wertevorrat uncertainties.
        For the probability of finding a v.b.s., we quote only the integration error.
        The preferred analysis is highlighted in gray.}
    \label{tab:sigmaResults}
    \begin{ruledtabular}
        \begin{tabular}{lcccclll}
            Prior  & $\sigma$ & $\ell_c/m_\pi$ & Bayes factor & Prob.\ of v.b.s. & \multicolumn{1}{c}{$E_0/m_\pi$}  & \multicolumn{1}{c}{$E_0/m_\pi$ (v.b.s.)} & \multicolumn{1}{c}{$E_0/m_\pi$ (res.)} \\ \hline
            Linear & 1.5      & 1.0            & 1.000        & 0.524(59)        & $1.874(10)(93) - i0.09(1)(11)$   & $1.878(11)(84)$      & $1.87(2)(10) - i0.192(12)(83)$         \\
            Adler  & 1.5      & 1.0            & 0.794        & 0.200(41)        & $1.882(11)(98) - i0.127(9)(82)$  & $1.79(3)(12)$        & $1.905(10)(74) - i0.158(9)(58)$         \\[\defaultaddspace]
            Linear & 1.5      & 1.5            & 1.000        & 0.786(48)        & $1.86(1)(11) - i0.05(1)(10)$     & $1.88(1)(10)$        & $1.79(3)(12) - i0.236(22)(84)$         \\
            Adler  & 1.5      & 1.5            & 0.639        & 0.333(50)        & $1.85(1)(10) - i0.12(1)(11)$     & $1.77(2)(12)$        & $1.883(10)(65) - i0.184(11)(82)$        \\[\defaultaddspace]
            \rowcolor{gray!25}[2\tabcolsep]
            Linear & 3.0      & 1.0            & 1.000        & 0.514(54)        & $1.883(11)(98) - i0.08(1)(10)$   & $1.87(1)(11)$        & $1.901(14)(87) - i0.163(14)(87)$        \\
            Adler  & 3.0      & 1.0            & 0.838        & 0.328(50)        & $1.89(1)(11) - i0.099(10)(90)$   & $1.84(2)(13)$        & $1.919(12)(82) - i0.147(10)(71)$        \\[\defaultaddspace]
            Linear & 3.0      & 1.5            & 1.000        & 0.700(56)        & $1.86(1)(11) - i0.08(3)(15)$     & $1.865(11)(98)$      & $1.86(4)(13) - i0.25(6)(19)$            \\
            Adler  & 3.0      & 1.5            & 0.958        & 0.285(45)        & $1.860(9)(89) - i0.12(1)(10)$    & $1.83(2)(13)$        & $1.873(8)(60) - i0.166(13)(84)$
        \end{tabular}
    \end{ruledtabular}
\end{table*}

\begin{figure}
    \centering
    \includegraphics[width=1\linewidth]{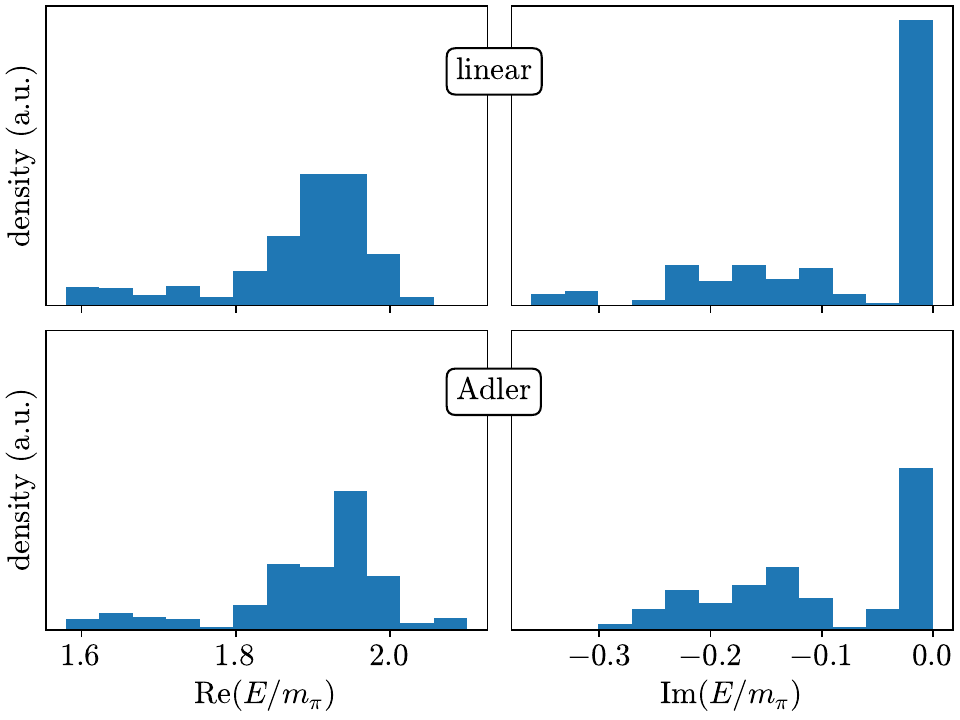}
    \caption{Histograms of the real (left) and imaginary (right) parts of the $\sigma$ pole locations for different priors.
        The upper panel shows the results obtained using a linear prior (\cf \cref{eq:priorLinear}) and the lower panel using an Adler prior (\cf \cref{eq:priorSigmaAdler}), both with $\sigma = 3$ and $\ell_c/m_\pi = 1$.
    }
    \label{fig:polePositionsHistogramsSigma}
\end{figure}

\Cref{fig:polePositionsHistogramsSigma} provides further visualization of our results regarding the $\sigma$ pole position, giving histograms of the real and imaginary parts of the pole location.
We remark that the rightmost bin for the imaginary part ends at zero rather than being centered there since we enforce $\imag E_0 \leq 0$ in the pole search in order to increase the stability of the solver. The rows of \cref{fig:polePositionsHistogramsSigma} also compare the dependence of the results on the linear versus Adler priors.
Both prior choices yield similar distributions for the real part of the pole position.
Moreover, the magnitudes of the imaginary part are similar when the pole is located away from the real line.
The primary difference is that the linear prior yields a pole on the real line (\ie, a virtual bound state) with somewhat greater probability (\cf \cref{tab:sigmaResults}).

The picture regarding pole position does not change significantly if more than one sample from the Wertevorrat is used on each configuration, indicating that one sample per configuration appears to suffice in this case to account for the uncertainty in the analytic continuation from a finite set of points.

\begin{figure}
    \centering
    \includegraphics[width=1\linewidth]{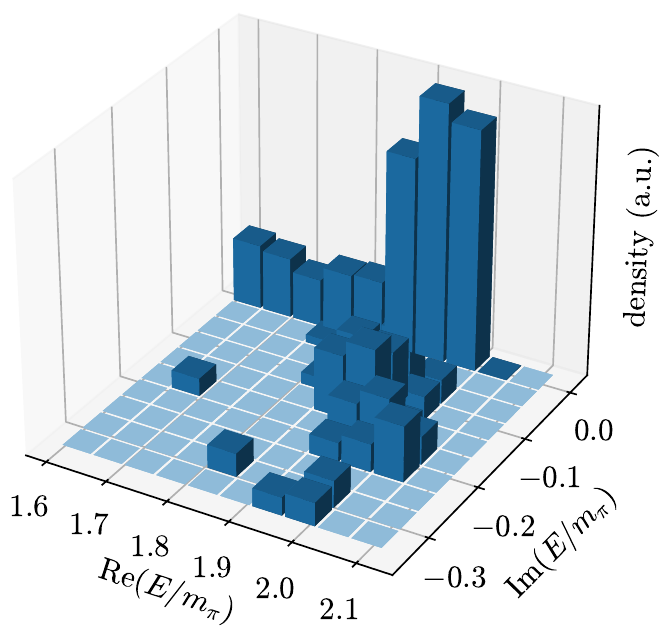}
    \caption{Two-dimensional histogram of the $\sigma$ pole location for our preferred prior (linear, $\sigma = 3$, and $\ell_c/m_\pi = 1$).
    Note that this shows the same information as the upper panel of \cref{fig:polePositionsHistogramsSigma}.
    }
    \label{fig:polePositionsHistogram2DSigma}
\end{figure}

As our final result, we choose to quote the analysis using the linear prior (as it has the larger Bayes factor) with $\sigma = 3$ and $\ell_c/m_\pi = 1$ (as these values introduce the mildest assumptions among the priors considered in \cref{tab:sigmaResults}).
This result corresponds to the upper panel in \cref{fig:polePositionsHistogramsSigma}
and is interpreted as a roughly 50\% probability for the $\sigma$ pole to be a virtual bound state at $m_\pi = 280~\mathrm{MeV}$.
A different representation of the result is given in \cref{fig:polePositionsHistogram2DSigma} as a full two-dimensional histogram.
The conclusions of this analysis compare well to the spread among different fit models considered in Ref.~\cite{Rodas:2023gma}.
However, the present analysis is able to provide a number that quantifies this uncertainty regarding the nature of the $\sigma$ pole.

\subsection{Coupled-channel \texorpdfstring{$\pi\Sigma - \bar{K}N$}{pi Sigma - Kbar N} scattering on D200}
\label{sec:L1405}

The present methods can also be applied to a coupled-channel system.
This section considers $I=0$ $\pi\Sigma - \bar{K}N$ scattering using data provided by the BaSc collaboration on the CLS ensemble D200 ($m_\pi \approx 200~\mathrm{MeV}$)~\cite{BaryonScatteringBaSc:2023zvt,BaryonScatteringBaSc:2023ori}.

This channel has been studied vigorously both in experiment and in theory for more than six decades, and open questions remain.
In 1959, a relatively narrow resonance was found experimentally just below the $\bar{K}N$ threshold \cite{Dalitz:1959dq,Dalitz:1959dn,Alston:1961zzd,Bastien:1961zz}, the $\Lambda(1405)$.
Nowadays, the status of the $\Lambda(1405)$ is listed with the highest (four-star) rating by the Particle Data Group (PDG) \cite{ParticleDataGroup:2024cfk}, indicating that ``existence is certain, and properties are at least fairly explored.''

However, the complete picture remains far from clear.
Unitarized chiral perturbation theory predicts the presence of another resonance pole just below the $\pi\Sigma$ threshold \cite{Oller:2000fj}, known as $\Lambda(1380)$.
Indeed, recent experimental reviews are at least compatible with the two-pole picture, but a definite conclusion has not been reached yet on the experimental side \cite{Anisovich:2020lec,Hyodo:2020czb,Mai:2020ltx}.
Accordingly, the $\Lambda(1380)$ is still only listed with a two-star rating in the PDG \cite{ParticleDataGroup:2024cfk}, indicating that ``evidence of existence is only fair.''

To elucidate the situation with lattice QCD,
Refs.~\cite{BaryonScatteringBaSc:2023zvt,BaryonScatteringBaSc:2023ori} presented the first
lattice-QCD calculation of the scattering amplitudes in this system using a coupled-channel analysis in the finite-volume formalism.
At a pion mass of about 200~MeV, they found clear evidence of two poles, a virtual bound state below the $\pi\Sigma$ threshold at $E_1/m_\pi = 6.856(45)$ and a resonance below the $\bar{K}N$ threshold at $E_2/m_\pi = 7.144(63) - i0.057(28)$.
Both poles were located on the Riemann sheet where ${\rm Im}\ q_{\pi\Sigma}<0$ and ${\rm Im} \ q_{\bar{K}N}>0$, which we will denote by $(-, +)$ in the following.

Here, we reanalyze the data from Ref.~\cite{BaryonScatteringBaSc:2023zvt,BaryonScatteringBaSc:2023ori} using our approach.
The goal is to provide more controlled estimates of model dependence in the amplitudes and pole positions.
Compared to the single-channel cases considered so far, the main difference is the need to parameterize three functions instead of one, corresponding to the three independent entries of the symmetric matrix $\tilde{\mathcal{K}}^{-1} = \eta 16\pi\sqrt{s}\mathcal{K}^{-1}$.
For a fixed number $n_p$ of energy nodes, the dimensionality of $\bm u$ increases by a factor three.
These three variables can be stacked into a single vector via
\begin{equation}
    \bm u = (\bm u^{(1)}, \bm u^{(2)}, \bm u^{(3)}),
    \label{eq:stacked_vector_u}
\end{equation}
where the elements correspond to the $\pi\Sigma; \pi\Sigma$, $\pi\Sigma; \bar{K}N$, and $\bar{K}N; \bar{K}N$ entries, respectively (\cf \cref{eq:QC:2channel}).

The present analysis uses a factorized prior of the form
\begin{align}
\begin{split}
    p(\bm u | \bm u_0) = \ & p_{\pi\Sigma; \pi\Sigma}(\bm u^{(1)} | \bm u^{(1)}_0) \times \\ &p_{\pi\Sigma; \bar{K}N}(\bm u^{(2)} | \bm u^{(2)}_0) \times \\&  p_{ \bar{K}N; \bar{K}N}(\bm u^{(3)} | \bm u^{(3)}_0).
    \label{eq:pfactor}
\end{split}
\end{align}
This factorized form could be generalized to include correlations between entries of the K matrix, but we currently find no compelling physical motivation for such correlation.
The priors for the diagonal elements, $p_{\pi\Sigma; \pi\Sigma}$ and $p_{ \bar{K}N; \bar{K}N}$, can be readily handled using Gaussian priors with covariance matrices
\begin{align}
\begin{split}
    &\Sigma^{(i)}_p( E,  E') = \\
    &\qquad \sigma^{(ii)}(E, E')^2 \exp\left[-\frac{(E - E')^2}{2\ell_c^2}\right] + \epsilon \delta_{E, E'},
    \label{eq:ppriorMultichannel}
\end{split}
\end{align}
analogous to the ones used in the single-channel analyses.
In contrast, the off-diagonal component must account for the fact that physical observables
(\eg pole positions or finite-volume energies) are invariant under flipping the sign of the off-diagonal $\pi\Sigma; \bar{K}N$ component of the K matrix~\cite{Oset:2012bf}. Thus, we use a prior that respects this symmetry,\footnote{Other choices of prior that also respect the symmetry are possible, \eg a log-normal or a folded-normal prior. }
\begin{align}
\begin{split}
    &p_{\pi\Sigma; \bar{K}N}(\bm u^{(2)} | \bm u^{(2)}_0) \propto \\
    &\ \exp\left[ -\frac{1}{2} (|\bm u^{(2)}| - \bm u^{(2)}_0)^T \left(\Sigma^{(2)}_{p}\right)^{-1} (|\bm u^{(2)}| - \bm u^{(2)}_0) \right] ,
\end{split}
\label{eq:priorProbabilityOffdiag}
\end{align}
where $|\bm u^{(2)}|$ indicates that the absolute value is taken for each component of $\bm u^{(2)}$, and $\Sigma^{(2)}_{p}$ is defined as in \cref{eq:ppriorMultichannel}. This prior is explicitly symmetric under sign flips of each component of $\bm u^{(2)}$, and the same is true for the posterior.

We consider a simple, constant prior defined by
\begin{equation}
    \bm u_0 = (0.1, \ 0, \ -0.1),
\end{equation}
where each component is understood to correspond to a vector with constant entries for each $\bm{u}^{(i)}$.
For the hyperparameters entering the prior covariance matrix, we compare different sets of choices to verify the stability of the results.
Using a similar compact notation for $\sigma^{(i)}$, we set
\begin{equation}
    \ \bm \sigma = (3.0, \ 0.8, \ 1.2) \times \tilde\sigma,
    \label{eq:priorWidthLambda1405}
\end{equation}
where $\tilde{\sigma} = 1$ or $\sqrt{2}$; and $\ell_c/m_\pi = 1.0$, 1.5, or 2.0. Note that $\ell_c$ is the same for all entries of the K matrix.
The scaling parameters for the prior width in \cref{eq:priorWidthLambda1405} have been chosen such that $\tilde{\sigma} = 0.1$
approximately corresponds to the typical error obtained from the best fit in Refs.~\cite{BaryonScatteringBaSc:2023zvt,BaryonScatteringBaSc:2023ori} (\cf also \cref{fig:KmatLambda1405} below).
Thus, we employ a prior width that is about one order of magnitude larger than the uncertainties of typical fits.

Compared to the single-channel case, a coupled-channel analysis hence introduces a considerably larger number of hyperparameters.
Also, the impossibility of visually inspecting the data for the variables of interest, \ie the entries of the K matrix, without performing any kind of analysis poses a significant challenge in obtaining guidance for these choices.
A major advantage of using our non-parametric approach in coupled-channel systems is that it circumvents the issue of introducing a large number of fit parameters that may not all be constrained by the data at hand, a common challenge observed in the fitting approach.

For the system studied in this subsection, we observe a breakdown in the sampling procedure based on the Gaussian approximation of the posterior which we used for sampling in all previously discussed cases (\cf \cref{sec:posteriorSampling}).
The breakdown manifests itself in very small effective sample sizes when trying to reweight to the exact posterior according to \cref{eq:reweightingImproved}.
Therefore, we instead use the Hybrid Monte Carlo (HMC) algorithm to sample the posterior robustly; the HMC algorithm is reviewed in \cref{app:HMC} for completeness.
A disadvantage of this approach is that, in addition to the significantly increased computing time required for the sampling, the Bayes factors cannot be computed exactly anymore.
\Cref{app:Bayes_factors_approx} explains how approximate Bayes factors can still be calculated, independent of the sampling procedure.
As we are only using a single choice of prior central values, we will not quote any Bayes factors in this section.
Since the sign of $\bm u^{(2)}$ is not physical and not fixed by the data, the component-wise absolute value is applied at the end of each HMC trajectory to avoid ambiguities in the interpolation of the points.
Following Refs.~\cite{BaryonScatteringBaSc:2023zvt,BaryonScatteringBaSc:2023ori}, the covariance matrix $\Sigma_\mathrm{d}$ used in \cref{eq:dataProbability} is the covariance matrix of the energy shifts with respect to the non-interacting levels.
Therefore, our results for the pole positions will be relative to the $\pi\Sigma$ threshold.

The analysis begins with a grid of 20 energy nodes equally spaced in $E/m_\pi \in [6.65, 7.85]$.
As above, the Schwarz--Christoffel rectangle mapping takes the complex energy plane (in pion-mass units) to the unit disk.
For finding the lower, virtual-bound-state pole, we employ a rectangle centered at $C = 7.25$, with full width $W = 1.8$ and full height $H = 0.1$, as the pole is expected to be on, or very close to, the real line.
For the upper, resonance pole, analytic continuation farther into the complex plane is required.
Here, the default rectangle is centered at $C = 7.25 - 0.075i$ and has height $H = 0.25$.
Depending on the prior, a second pole is found within this mapping region on 62\%--72\% of samples.
On the remaining samples, the mapping domain is shifted farther into the complex plane using $C = 7.25 - 0.1i$ and $H = 0.3$.

This procedure identifies a pole around the $\pi\Sigma$ threshold on all samples and another pole around the $\bar{K}N$ threshold on $\gtrsim 99.5\%$ of samples.
The tiny fraction of samples which do not feature a second pole are excluded from the average for this pole.
With this caveat, results for the second pole position are to be interpreted as conditional expectation values, conditioned on the existence of the second pole.

On at most 0.5\% of the samples (depending on the prior), the lower pole is found on the physical $(+, +)$ sheet, corresponding to a bound state, instead of the $(-, +)$ sheet.
We remark that the original analysis in Refs.~\cite{BaryonScatteringBaSc:2023zvt, BaryonScatteringBaSc:2023ori} also found the lower pole to be bound on roughly 0.5\% of the bootstrap samples.

The imaginary part of the upper pole is smaller (in magnitude) than 0.005 for 4\%--9\% of the samples in the present analysis.
On most of these samples, an additional pole close to the real line is observed on the physical sheet.
In the approximation where the two scattering channels decouple, this phenomenon can be interpreted as the two complex conjugate resonance poles converging towards the real line.
One of these poles transitions to the physical sheet, while the other remains on the unphysical sheet.

Furthermore, about 10\%--20\% of our samples feature additional pole(s) between the $\pi\Sigma$ and $\bar{K}N$ thresholds on the $(-, +)$ sheet.
For around 2\%--5\%, we find, employing any of the above mappings, poles in the complex plane on the physical $(+, +)$ sheet, which violates causality.\footnote{It is justified to exclude the corresponding samples from the average in order to enforce causality. However, since this observation is not statistically significant, we ignore it in the present analysis.}
The occurrence of such additional poles, no matter on which sheet, depends in nearly all cases on the details of the chosen mapping.
Hence, we strongly suspect them to be interpolation artifacts which could be eliminated by employing a much finer grid.
In fact, by employing a grid of 24 (rather than 20) equally spaced energy nodes in the aforementioned interval, we find that for the prior with $\tilde{\sigma} = \sqrt{2}$ and $\ell_c/m_\pi = 1$, the fraction of samples with causality-violating poles decreases from 2\% to 0.5\%.
We remark that unphysical poles were also observed in the fitting approach in Refs.~\cite{BaryonScatteringBaSc:2023zvt,BaryonScatteringBaSc:2023ori}, and are ultimately related to the statistical uncertainty of the lattice-QCD energy levels.
In the following, we only quote the two pole positions which are stable under variation of the mappings and which consistently occur on (almost) all samples.

\Cref{tab:Lambda1405Results} collects results for the two pole positions which appear consistently and the ratio of their residues between the two channels.
We observe some (typically few-percent) outliers in the distribution of poles, reflecting the significantly more complicated analytic structure for the scattering amplitudes in a coupled-channel system.
We therefore employ the median and the central 68\% quantiles as outlier-robust replacements for the mean and standard deviation, respectively.
This choice also readily accommodates asymmetric errors in cases where the distributions are significantly non-Gaussian.

\Cref{tab:Lambda1405Results} shows that results from different priors are consistent within the quoted uncertainties.
Moreover, the size of the errors changes little under variation of the hyperparameters $\tilde{\sigma}$ and $\ell_c$.
This stability indicates that our chosen values are well within the regime where the results are dominated by the data rather than by the prior.
Our values are consistent at the level of 1--2 standard deviations with the fit-based original analysis in Refs.~\cite{BaryonScatteringBaSc:2023zvt,BaryonScatteringBaSc:2023ori}, with larger errors.

\begingroup
\renewcommand{\arraystretch}{1.5}
\begin{table*}[htb]
    \caption{Results for the coupled-channel $\pi\Sigma - \bar{K}N$ system with different priors.
        For the pole positions and residues, the first error refers to the Monte-Carlo integration error and the second one to the combined Bayes and Wertevorrat uncertainties.
        Asymmetric errors are always on the numbers including the sign, \ie the lower error is going \emph{away} from zero.
        The preferred analysis is highlighted in gray.}
    \label{tab:Lambda1405Results}
    \begin{ruledtabular}
        \begin{tabular}{lcccllll}
            Prior  & $\tilde{\sigma}$ & $\ell_c/m_\pi$ & \multicolumn{1}{c}{$(E_1 - m_\pi - m_\Sigma)/m_\pi$} & \multicolumn{1}{c}{$|g^{(1)}_{\pi\Sigma} / g^{(1)}_{\bar{K}N}|$} & \multicolumn{1}{c}{$(E_2 - m_\pi - m_\Sigma)/m_\pi$}       & \multicolumn{1}{c}{$|g^{(2)}_{\pi\Sigma} / g^{(2)}_{\bar{K}N}|$} \\ \hline
            const. & 1                & 1.0            & $-0.034(2)_{-0.025}^{+0.014}$                        & $2.5(3)_{-1.1}^{+3.3}$                                           & $0.309(4)_{-0.034}^{+0.047} - i0.041(4)_{-0.067}^{+0.031}$ & $0.48(2)_{-0.20}^{+0.21}$ \\
            const. & 1                & 1.5            & $-0.031(2)_{-0.034}^{+0.013}$                        & $2.2(3)_{-0.9}^{+1.7}$                                           & $0.303(4)_{-0.041}^{+0.058} - i0.042(4)_{-0.071}^{+0.027}$ & $0.49(2)_{-0.18}^{+0.19}$ \\
            const. & 1                & 2.0            & $-0.028(1)_{-0.027}^{+0.008}$                        & $1.9(1)_{-0.7}^{+1.0}$                                           & $0.306(4)_{-0.032}^{+0.056} - i0.048(4)_{-0.051}^{+0.031}$ & $0.51(2)_{-0.19}^{+0.18}$ \\[\defaultaddspace]
            \rowcolor{gray!25}[\tabcolsep][11\tabcolsep]
            const. & $\sqrt{2}$       & 1.0            & $-0.031(2)_{-0.033}^{+0.015}$                        & $2.6(3)_{-1.1}^{+2.1}$                                           & $0.318(4)_{-0.039}^{+0.048} - i0.032(3)_{-0.068}^{+0.023}$ & $0.50(2)_{-0.20}^{+0.20}$ \\
            const. & $\sqrt{2}$       & 1.5            & $-0.032(2)_{-0.038}^{+0.011}$                        & $2.3(2)_{-0.9}^{+1.8}$                                           & $0.308(4)_{-0.033}^{+0.046} - i0.036(3)_{-0.056}^{+0.026}$ & $0.47(2)_{-0.19}^{+0.22}$ \\
            const. & $\sqrt{2}$       & 2.0            & $-0.033(2)_{-0.028}^{+0.011}$                        & $2.1(4)_{-0.8}^{+1.6}$                                           & $0.307(4)_{-0.038}^{+0.063} - i0.044(4)_{-0.081}^{+0.030}$ & $0.47(1)_{-0.18}^{+0.22}$
        \end{tabular}
    \end{ruledtabular}
\end{table*}
\endgroup

As above, the preferred analysis conservatively chooses the loosest prior with $\tilde{\sigma} = \sqrt{2}$ and $\ell_c/m_\pi = 1$.
\Cref{fig:KmatLambda1405} shows the posteriors for the three entries of $\tilde{\mathcal{K}}^{-1}$ along the real axis together with the prior;
the figure also shows the agreement with the best-fit results from Refs.~\cite{BaryonScatteringBaSc:2023zvt,BaryonScatteringBaSc:2023ori}.
As expected, the overall errors in the present analysis show a modest increase compared to those in Refs.~\cite{BaryonScatteringBaSc:2023zvt,BaryonScatteringBaSc:2023ori}, since the present analysis gives an error band representing the combined statistical plus systematic uncertainty.

All variations of our analysis suggest a slight preference for
$\tilde{\mathcal{K}}^{-1}_{\bar{K}N; \bar{K}N}$ to be positive at the lower end of the energy interval, while the best-fit result from Refs.~\cite{BaryonScatteringBaSc:2023zvt,BaryonScatteringBaSc:2023ori} preferred a small negative value.
This difference illustrates the limitations of choosing a particular parametric form for the fit function;
the choice employed in Refs.~\cite{BaryonScatteringBaSc:2023zvt,BaryonScatteringBaSc:2023ori} enforced $\tilde{\mathcal{K}}^{-1}_{\bar{K}N; \bar{K}N}$ to have the same sign at all energies.
In this sense, the proposed Bayesian analysis offers greater flexibility and naturally allows for a sign change here. While this could in principle also be implemented in the fitting approach, complicated fit forms can lead to overfitting, which is not a concern in the Bayesian framework.

\begin{figure}
    \centering
    \includegraphics[width=1\linewidth]{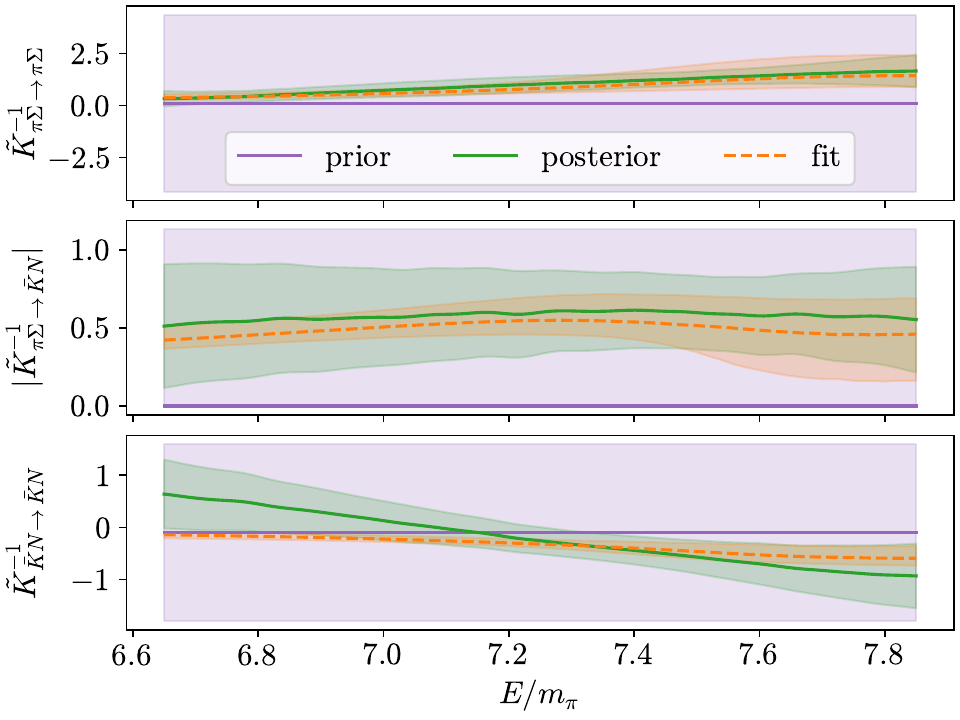}
    \caption{Prior (purple) and posterior (green) for $\tilde{\mathcal{K}}^{-1}$ in the coupled-channel $\pi\Sigma - \bar{K}N$ system, using $\tilde{\sigma} = \sqrt{2}$ and $\ell_c/m_\pi = 1$.
        The green band shows the Bayes uncertainty only, where the Monte-Carlo integration error and the size of the Wertevorrat are negligibly small.
        The dashed orange lines and bands depict the result from the best fit of Refs.~\cite{BaryonScatteringBaSc:2023zvt,BaryonScatteringBaSc:2023ori} and its statistical uncertainty.
        In all cases, the lines and bands refer to the median and quantile errors, respectively.
    }
    \label{fig:KmatLambda1405}
\end{figure}

\begin{figure*}
    \centering
    \subfloat{\includegraphics[width=0.5\linewidth]{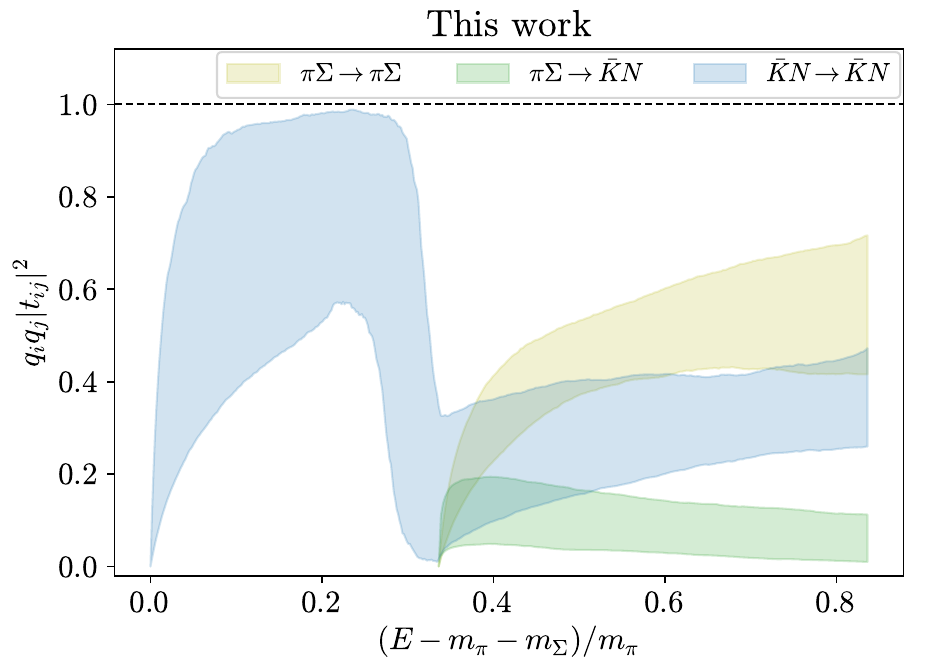}}
    \subfloat{\includegraphics[width=0.5\linewidth]{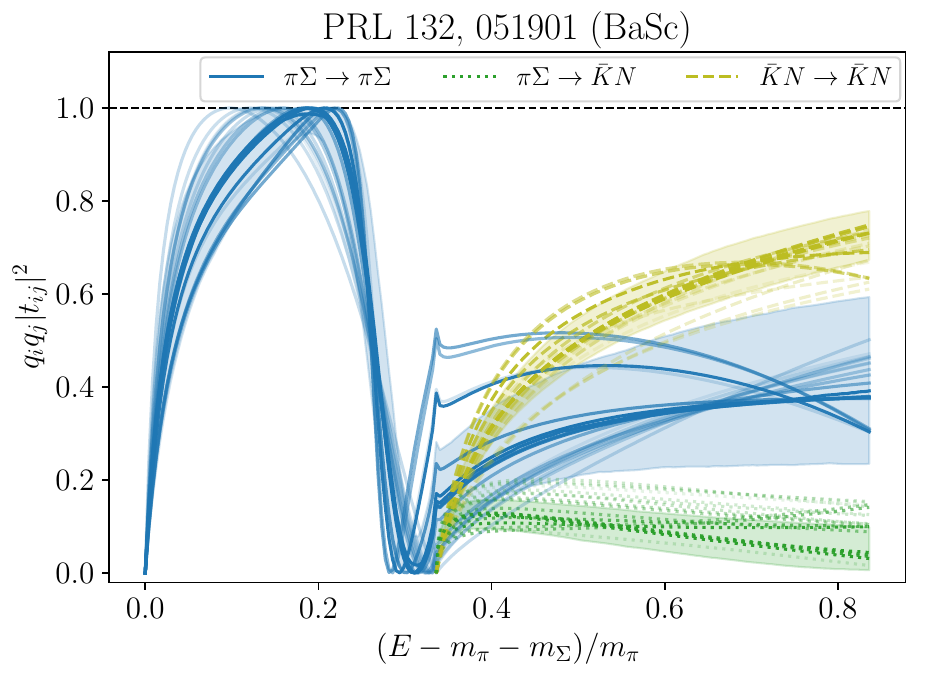}}
    \caption{Posteriors for $q_i q_j |t_{ij}|^2$ for the coupled-channel $\pi\Sigma - \bar{K}N$ system (left panel).
        We use the prior with $\tilde{\sigma} = \sqrt{2}$ and $\ell_c/m_\pi = 1$.
        In the right panel, we compare to the analysis of Refs.~\cite{BaryonScatteringBaSc:2023zvt,BaryonScatteringBaSc:2023ori}.
        Here, the band depicts the statistical uncertainty of the best fit, while the collection of lines shows the model variations.
        The transparency of each line is proportional to $\exp[-(\mathrm{AIC} - \mathrm{AIC}_\mathrm{min}) / 2]$, where $\mathrm{AIC}_\mathrm{min}$ is the lowest AIC, corresponding to the band.
        In both panels, the bands refer to the quantile errors.}
    \label{fig:scatteringAmplitudeLambda1405}
\end{figure*}

\Cref{fig:scatteringAmplitudeLambda1405} presents the same results in terms of ${t^{-1} = \eta 16 \pi \sqrt{s} \mathcal{M}^{-1}}$, which mirrors the presentation in Refs.~\cite{BaryonScatteringBaSc:2023zvt,BaryonScatteringBaSc:2023ori} (see also Ref.~\cite{Dudek:2016cru}).
In this parameterization, each component fulfills $0 \leq q_i q_j |t_{ij}|^2 \leq 1$ in the energy region where both the $i$ and $j$ channels are open.
The left panel of \cref{fig:scatteringAmplitudeLambda1405} shows the result of the present analysis, while the right panel summarizes the results of Ref.~\cite{BaryonScatteringBaSc:2023ori}.
On the left, the error bands in the present analysis naturally incorporate both the statistical error from the Monte-Carlo sampling of the gauge field and the parameterization dependence of the K matrix.
In Ref.~\cite{BaryonScatteringBaSc:2023ori}, the results contained in the right panel were presented in two separate figures (Figs.~6 and 7 therein), one showing the best-fit statistical uncertainty and one estimating the systematic uncertainty from model dependence.
Good qualitative agreement is observed between the Bayesian approach and the previous analysis based on explicit parametric fits.
The fact that the statistical and systematic uncertainties can be combined into one total error band is a major advantage of our procedure.

\begin{figure}
    \centering
    \includegraphics[width=1\linewidth]{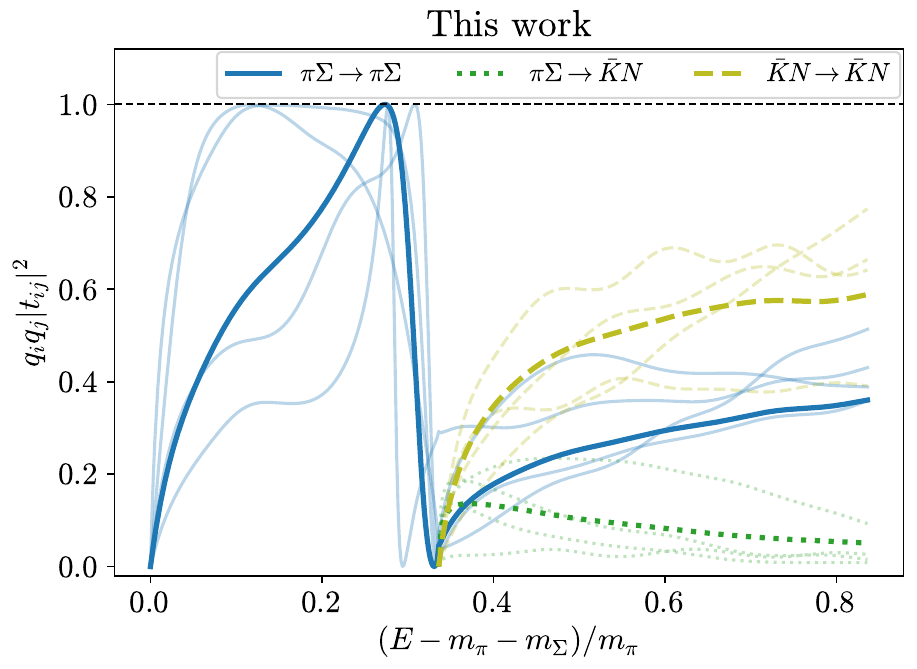}
    \caption{Samples from the posterior distribution for $q_i q_j |t_{ij}|^2$ for the coupled-channel $\pi\Sigma - \bar{K}N$ system.
             The transparent lines indicate four randomly selected samples from the full set we have generated.
             The thick, opaque lines show the central value, which is calculated by taking the average of all samples at the level of $\tilde{\mathcal{K}}^{-1}$ and then computing the quantity $q_i q_j |t_{ij}|^2$ from this sample mean.}
    \label{fig:scatteringAmplitudeLambda1405Samples}
\end{figure}

\Cref{fig:scatteringAmplitudeLambda1405} (left) displays the central 68\% quantiles for $q_i q_j |t_{ij}|^2$ at each energy.
To illustrate the typical behavior as a function of energy, we show in \cref{fig:scatteringAmplitudeLambda1405Samples} four randomly selected samples from the posterior distribution, along with the central value.
It is important to note that in order to extract the correct functional behavior, the central value must be taken at the level of $\tilde{\mathcal{K}}^{-1}$, and only afterwards converted to $q_i q_j |t_{ij}|^2$.
Given the considerable variability observed in the location of the peak in the $\pi\Sigma \to \pi\Sigma$ component and the narrow nature of the peak on many samples, as illustrated in \cref{fig:scatteringAmplitudeLambda1405Samples}, taking the mean over samples for $q_i q_j |t_{ij}|^2$ would effectively wash out any peak structure.
The same holds for the cusp at the opening of the $\bar{K}N$ threshold, which is observed on all samples (though it is occasionally very small), but is washed out in the error bands shown in \cref{fig:scatteringAmplitudeLambda1405} (left).

The preferred results for the pole locations (taken from the prior with $\sigma = \sqrt{2}$ and $\ell_c/m_\pi = 1$) are
\begin{align}
    \label{eq:pole1Lambda1405}
    E_1/m_\pi &= 6.832(2)_\mathrm{MC}(35)_\mathrm{thr}{}_{-0.033}^{+0.015} , \\
    \label{eq:couplingRatio1Lambda1405}
    |g^{(1)}_{\pi\Sigma} / g^{(1)}_{\bar{K}N}| &= 2.6(3)_\mathrm{MC}{}_{-1.1}^{+2.1} , \\
    \label{eq:pole2Lambda1405}
    \begin{split}
    E_2/m_\pi &= 7.182(4)_\mathrm{MC}(35)_\mathrm{thr}{}_{-0.039}^{+0.048} \\
    &\quad {} - i0.032(3)_\mathrm{MC}{}_{-0.068}^{+0.023} ,
    \end{split} \\
    \label{eq:couplingRatio2Lambda1405}
    |g^{(2)}_{\pi\Sigma} / g^{(2)}_{\bar{K}N}| &= 0.50(2)_\mathrm{MC}{}_{-0.20}^{+0.20} .
\end{align}
The first error refers to the Monte-Carlo integration error; the second error (which is only present for the real part of the pole positions) originates from the error of the two-particle threshold $(m_\Sigma + m_\pi)/m_\pi$; and the third, asymmetric error is the combined Bayes and Wertevorrat uncertainty.
The asymmetric errors on $\imag E_2/m_\pi$ are on the number including the sign, \ie the lower error is going \emph{away} from zero.

\begin{figure}
    \centering
    \includegraphics[width=1\linewidth]{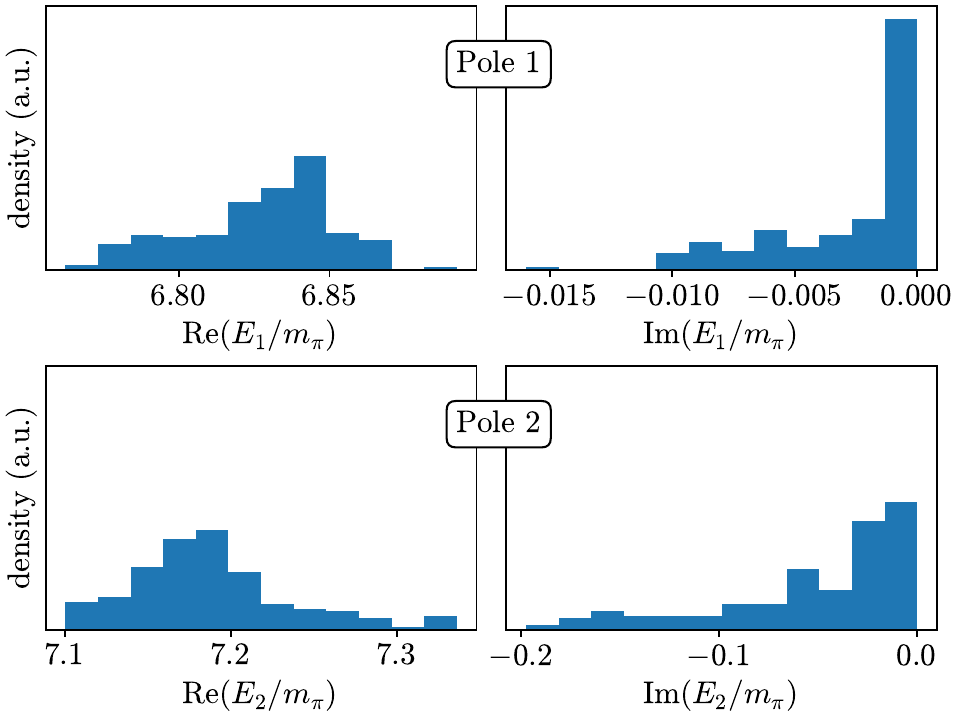}
    \caption{Histograms of the real (left) and imaginary (right) parts of the two pole locations in the $\pi\Sigma - \bar{K}N$ system.
        We use the prior with $\tilde{\sigma} = \sqrt{2}$ and $\ell_c/m_\pi = 1$.
    }
    \label{fig:polePositionsHistogramsLambda1405}
\end{figure}

\Cref{fig:polePositionsHistogramsLambda1405} shows the histograms of the real and imaginary parts of the two poles.
As can be seen from the top right histogram, $\imag E_1/m_\pi$ is concentrated at (or close to) zero, with only one sample clearly featuring a resonance above the $\pi \Sigma$ threshold.
For the remaining samples, the small imaginary part can be explained by the Wertevorrat uncertainty.
Consequently, $\imag E_1/m_\pi$ is compatible with zero within the combined Bayes and Wertevorrat uncertainty.
We conclude that the lower pole is most likely a virtual bound state, as found in Refs.~\cite{BaryonScatteringBaSc:2023zvt,BaryonScatteringBaSc:2023ori}, and choose not to quote an imaginary part on it.

\section{Conclusion}
\label{sec:conclusion}

This work presents a novel approach to computing scattering amplitudes in the finite-volume formalism using lattice QCD
by combining Bayesian inference with analytic continuation techniques. Specifically, by treating the constraints on scattering amplitudes as an inverse problem, we draw inspiration from methods used in spectral-function reconstruction, such as Gaussian processes~\cite{DelDebbio:2021whr,Pawlowski:2022zhh,Horak:2023xfb,DelDebbio:2024lwm,Dutrieux:2024rem,Dutrieux:2025jed} and Nevanlinna interpolation~\cite{PhysRevLett.126.056402,Bergamaschi:2023xzx}.
The methodology presented here proceeds in two steps: the solution of constraints on the real axis using Bayesian inference followed by numerical analytic continuation to the complex plane to determine resonance properties.

The first step solves the finite-volume quantization condition to
constrain the shape of the inverse K matrix at real-valued energies.
The starting point is a Bayesian prior informed by existing knowledge, such as effective field theories or previous lattice-QCD results.
The prior includes hyperparameters---primarily a ``width'', which encodes the uncertainty in prior knowledge, and a ``correlation length'', which controls the smoothness or variability of the function across energy. Bayes' theorem then furnishes samples for the inverse K matrix at a discrete set of nodes.
The resulting posterior distribution provides both central values and correlated uncertainties.

A key feature of this approach is that it does not rely on any specific functional form for the inverse K matrix, making the method non-parametric.
While the result in principle still depends on the chosen prior hyperparameters, this dependence is much milder than the restriction to specific functional forms in the fitting approach.
In the Bayesian approach, all possible smooth functions contribute to the posterior, each with its corresponding weight.
Samples drawn from the posterior can thus be interpreted as functional samples of the inverse K matrix.
In this way, both statistical uncertainties and parametrization dependence are naturally encoded in the width of the posterior distribution.

The second step is the numerical analytic continuation from a finite set of points.
Each functional sample from the posterior defines a possible realization of the inverse K matrix in terms of its values at a discrete set of energies.
Nevanlinna--Pick interpolation provides the key tool for this step: given a discrete set of nodes and their corresponding values, we map them onto the unit disc in both the domain and codomain, and solve the interpolation problem in that space. This approach performs the analytic continuation on a simply connected open subset of the complex plane and assumes that the function (here, the inverse K matrix) is analytic within that region. We emphasize that, for the rigorous bounds to apply, the conformal maps which transform the problem to the unit disk must correctly bound the codomain of the inverse scattering amplitude.

Since only a discrete set of points is known, the interpolation problem admits many possible solutions.
The set of all such solutions is characterized in Nevanlinna--Pick interpolation by the \emph{Wertevorrat}.
By increasing the number of nodes in the Bayesian analysis, the size of the Wertevorrat can be made arbitrarily small (see, \eg, Ref.~\cite{Jay:2025dzl}).
For a finite set of nodes, the associated uncertainty can be combined with the statistical uncertainty by sampling uniformly within the Wertevorrat.

We have presented several numerical examples to illustrate the methodology. First, in \cref{sec:toy_example}, we consider a toy model with Gaussian noise, for which the exact solution is known. This model mimics the behavior of the \( \rho \) resonance in QCD around \( m_\pi = 280 \) MeV. We demonstrate the application of the method and show that the resulting phase shift and pole position agree with the exact solution within the quoted uncertainties.

We also applied the methodology to three examples using real lattice-QCD data. The first example concerns the \( I = 2 \) \( \pi\pi \) system from Refs.~\cite{Blanton:2021llb,Dawid:2025doq}.
In this case, only the analysis for real energies is performed, as there are no resonances present.
We extract the scattering length and effective range, including parametrization dependence, thereby improving the analysis and uncertainty quantification of Ref.~\cite{Dawid:2025doq}.

The remaining two QCD examples involve hadronic resonances: the \( \sigma \) resonance~\cite{Briceno:2016mjc,Rodas:2023gma} and the double-pole structure associated with the \( \Lambda(1405) \) resonance~\cite{BaryonScatteringBaSc:2023ori,BaryonScatteringBaSc:2023zvt}.

For the \( \sigma \) resonance, we build upon previous analyses to provide a quantitative estimate of the probability of finding either a virtual bound state or a resonance for the $\sigma$ at $m_\pi \approx  280$ MeV.
We find that both scenarios are approximately equally likely given the current dataset.
The extracted pole locations are consistent with those reported by the HadSpec Collaboration across the different models they tested, but our analysis provides an uncertainty that includes parametrization dependence. It is noteworthy that in the special case of low-energy $\pi\pi$ scattering, a different method to reduce parameterization dependence was proposed in Ref.~\cite{Rodas:2023nec}, based on dispersion relations that implement crossing symmetry of the scattering amplitudes. That method is complementary in the sense that it actually reduces the parameterization dependence, while the one presented here aims to quantify it more robustly. The two methods could plausibly be combined in order to achieve a non-parametric analysis that also enforces crossing symmetry.

Finally, in the \( \Lambda(1405) \) case, we demonstrate the application of our method to coupled-channel scattering. While more computationally demanding, the analysis follows the same general procedure. A mild complication arises from enforcing the symmetries of the K matrix, particularly in its off-diagonal elements. Our results are consistent with those of Refs.~\cite{BaryonScatteringBaSc:2023ori,BaryonScatteringBaSc:2023zvt}, but we also improve the uncertainty quantification by incorporating parametrization dependence into the overall uncertainty.

The proposed method represents a significant step forward in providing reliable estimates of uncertainties in the extraction of scattering amplitudes and resonance properties from lattice QCD.
The numerical demonstrations presented here show that a non-parametric approach enables more robust uncertainty quantification. Similar ideas may be applicable to systems involving three hadrons, where a short-range three-body interaction must be parametrized~\cite{Blanton:2019vdk,Hansen:2020otl,Blanton:2021llb,Garofalo:2022pux,Brett:2021wyd,Mai:2021nul,Draper:2023boj,Yan:2024gwp}.
Another promising future direction lies in exploring synergies between the present work and analytic continuation methods used in phenomenological studies of resonances~\cite{Pelaez:2022qby}, which could lead to improved constraints on resonance properties across both lattice QCD and experimental data.

\section*{Acknowledgements}

Finite-volume energy levels taken from Ref.~\cite{Rodas:2023gma} were provided by the Hadron Spectrum Collaboration---no endorsement on their part of the analysis presented in the current paper should be assumed.
We would also like to warmly thank the Baryon Scattering Collaboration~\cite{BaryonScatteringBaSc:2023ori,BaryonScatteringBaSc:2023zvt} and the authors of Ref.~\cite{Blanton:2021llb} for providing their data for the finite-volume energy levels.

We acknowledge helpful discussions with Arnau Beltran Martínez, John Bulava, Arkaitz Rodas, Alexander Segner, Steve Sharpe, Christopher Thomas, and Julian Urban.

WJ gratefully acknowledges kind hospitality from the ECT* in Trento, Italy during the completion of this work together with useful discussions at the workshop ``The complex structure of strong interactions in Euclidean and Minkowski space.''

Part of the calculations were performed on UBELIX (\url{https://www.id.unibe.ch/hpc}), the HPC cluster at the University of Bern.

\appendix

\section{HMC for sampling the posterior}
\label{app:HMC}

This appendix discusses the implementation of the Hybrid Monte Carlo (HMC) algorithm~\cite{Duane:1987de} to sample the posterior, which is used in this work for the analysis of the $\Lambda(1405)$ case.

The posterior distribution to be sampled is $ p(\bm{u} | \bm{E}_\mathrm{d}, \bm{u}_0)$, which we abbreviate in this appendix as $p(\bm{u})$. Thus, the distribution has a dimensionality of $n_p n_c$ , where $n_p$ denotes the number of nodes, and $n_c$ denotes the number of elements in the K matrix that need to be parametrized; $n_c=3$ in the case of the $\Lambda(1405)$.
For this algorithm, it is conventional to define the action as $S(\bm u) = - \log p(\bm{u})$.

We use the standard HMC algorithm~\cite{Duane:1987de} defined by the Hamiltonian
\begin{equation}
    H(\bm{u}, \bm{\pi}) = S(\bm u) + \frac{1}{2} \bm{\pi}^2,
\end{equation}
where $\bm{\pi}$ are the conjugate Gaussian momenta, which have the same dimensionality as the variable $\bm u$.

The equations of motion of Hamiltonian dynamics are solved to generate samples:
\begin{align}
    \frac{d\bm{u}}{d \tau } =  \bm \pi , \quad \quad
     \frac{d\bm{\pi}}{d \tau } =  - \bm F(\bm u),
\end{align}
where $\bm F$ is the force, and each of its components is calculated as
\begin{equation}
    F_i = \frac{d S(\bm u)}{d u_i}. \label{eq:hmcforce}
\end{equation}

The algorithm proceeds as follows. A sample of momenta is generated at the beginning of each trajectory, and the equations of motion are solved numerically up to $\tau = \tau_\mathrm{HMC}=1$. As is standard in lattice QCD, symplectic integrators are used~\cite{Gattringer:2010zz,Omelyan:2002qkh,Takaishi:1999bi}. To correct for inexactness of the numerical integration, the HMC algorithm includes an accept-reject step.

We now describe some technical details of our implementation. First, in order to perform the HMC, we use simple (linear) spline interpolations, rather than Nevanlinna to define the action $S(\bm u)$. This is faster and more stable for the Hamiltonian evolution, but leads to no visible change in the K matrices on the real axis.  In addition, we compute the force in \cref{eq:hmcforce} numerically using finite differences. This step involves computing the variation of the energy levels predicted by the quantization condition with respect to the parameters $\bm u$.

In order to accelerate the generation of samples, we use a two-level integrator, inspired by the Hasenbusch preconditioning in lattice QCD~\cite{Hasenbusch:2001ne,Hasenbusch:2002ai}. Specifically, we split the action as:
\begin{equation}
     S(\bm u)  =  \Delta S(\bm u) +  S_{\rm approx}(\bm u),
\end{equation}
where $ S_{\rm approx}$ is an approximate action whose forces are fast to compute. The correction $\Delta S =S- S_{\rm approx} $ is smaller in magnitude, but contains the L\"uscher equation such that the computation of forces is slower. In the outer level, we compute $\Delta S$ and use a 4th-order Omelyan integrator~\cite{Omelyan:2002qkh,Takaishi:2005tz} with 20--40 steps, while in the inner integrator we use a simple Leapfrog integrator~\cite{Gattringer:2010zz} with 40 steps. In this work, we use an approximate action obtained by approximating the L\"uscher equation as a second-order polynomial:
\begin{align}
\begin{split}
    E(\bm u)_i = E_{L,i} &+ J_{ij} (\bm u - \bm u_L)_j \\
    & {} + \frac{1}{2} H_{ijk} (\bm u - \bm u_L)_j (\bm u - \bm u_L)_k ,
\end{split}
\end{align}
where the subscript $L$ denotes the expansion point as in \cref{eq:linearizationQC}.
In this form, we typically achieve acceptance rates of about 80\%--90\%.

\section{Approximate Bayes factors}
\label{app:Bayes_factors_approx}
In case the posterior distribution differs significantly from the approximate distribution, so that the reweighting approach breaks down, the Bayes factors cannot be computed as described in \cref{sec:prior_dependence} any longer.
The obvious alternative, Markov-Chain Monte Carlo methods such as HMC, do not allow for a straightforward extraction of the normalizing constant of the probability distribution.
Hence, in such situations, only an approximate Bayes factor obtained from the Gaussian approximation of the posterior distribution can be straightforwardly computed.
Linearizing the quantization condition according to \cref{eq:linearizationQC} and noting that $\int d^{n_p}u \, p^*(\bm u) = 1$, gives
\begin{equation}
    Z_\mathrm{post, approx} = \frac{\exp\left[ -\frac{1}{2} \Delta \bm E^T (\Phi_d^*)^{-1} \Delta \bm E \right] }{\sqrt{(2\pi)^{n_p} \det \Phi_d^*}} \,
    \label{eq:Zpost_approx}
\end{equation}
where
\begin{align}
    \Phi_d^* &= \Sigma_d + J \Sigma_p J^T , \\
    \Delta \bm E &= \bm E_\mathrm{d} - \bm E_L - J (\bm u_0 - \bm u_L) .
\end{align}

\end{document}